\documentclass[12pt,a4paper]{article}
\pdfoutput=1
\usepackage{jheppub}
\usepackage{dsfont}
\usepackage{amssymb,amsmath}
\usepackage{epsfig}
\usepackage{epstopdf}
\usepackage{latexsym}
\usepackage{graphicx}
\usepackage{subfigure}
\usepackage{booktabs}
\usepackage{bbm}
\pdfoutput=1
\makeatletter
\def\@fpheader{\relax}
\makeatother
\newcommand{\be}{\begin{equation}}
\newcommand{\ee}{\end{equation}}
\newcommand{\bea}{\begin{eqnarray}}
\newcommand{\eea}{\end{eqnarray}}

\def\cL{{\cal L}}

\def\cL{{\cal L}}

\subheader{\begin{flushright}
UTTG-08-13\\
TCC-005-13
\end{flushright}}

\title{Strong Subadditivity, Null Energy Condition and Charged Black Holes}

\author[a,b]{Elena Caceres,}
\author[b]{Arnab Kundu,}
\author[b,c]{Juan F. Pedraza}
\author[b,c]{and Walter Tangarife}

\affiliation[a]{Facultad de Ciencias, Universidad de Colima, Bernal Diaz del Castillo 340, Colima, Mexico}
\affiliation[b]{Theory Group, Department of Physics, The University of Texas, Austin, TX 78712, USA}
\affiliation[c]{Texas Cosmology Center, The University of Texas, Austin, TX 78712, USA}

\emailAdd{elenac@zippy.ph.utexas.edu}
\emailAdd{arnab@physics.utexas.edu}
\emailAdd{jpedraza@physics.utexas.edu}
\emailAdd{wtang@physics.utexas.edu}

\abstract{Using the Hubeny-Rangamani-Takayanagi (HRT) conjectured formula for entanglement entropy in the context of the AdS/CFT correspondence with time-dependent backgrounds, we investigate the relation between the bulk null energy condition (NEC) of the stress-energy tensor with the strong sub-additivity (SSA) property of entanglement entropy in the boundary theory. In a background that interpolates between an AdS to an AdS--Reissner-Nordstrom-type geometry, we find that generically there always exists a critical surface beyond which the violation of NEC would naively occur. However, the extremal area surfaces that determine the entanglement entropy for the boundary theory, can penetrate into this {\it forbidden region} only for certain choices for the mass and the charge functions in the background. This penetration is then perceived as the violation of SSA in the boundary theory. We also find that this happens only when the critical surface lies above the apparent horizon, but not otherwise. We conjecture that SSA, which is thus non-trivially related to NEC, also characterizes the entire time-evolution process along which the dual field theory may thermalize.}

\begin{document}

\maketitle
\flushbottom

\section{Introduction\label{Intro}}

Black holes have become, to a modern day's theoretical physicist, an instructive toy to play with: the ``harmonic oscillator" {\it \`{a} la mode}. The very nature of black hole entropy, which states that the number of degrees of freedom in a theory of quantum gravity scales as the area, gave birth to the idea of holography\cite{Susskind:1994vu, Banks:1996vh}. Subsequently, a concrete realization of the holographic principle was conceived under the name of the AdS/CFT correspondence\cite{Maldacena:1997re}. This correspondence, a hitherto {\it unproven}\footnote{It is often debated what a {\it proof} might mean in this context. In fact, it has been suggested that this correspondence should perhaps be viewed as a definition of quantum gravity in AdS-space and not as a proposition that is amenable to proof.} but compelling conjecture, has also emerged to be a remarkable tool in addressing aspects of strongly coupled large $N$ gauge theories.

Within this context, it is possible to realize ideas that are natural in quantum information theory: one such example is entanglement entropy. Entanglement entropy, defined as the von-Neumann entropy with respect to a reduced density matrix, measures the quantum entanglement of a system, and thus becomes an interesting quantity to analyze specially for systems described by strongly coupled quantum field theories. For large $N$ gauge theories, whose gravity duals are described by Einstein gravity (with a negative cosmological constant) in the presence of a suitable matter field, entanglement entropy can be computed using the Ryu-Takayanagi (RT) conjectured formula\cite{Ryu:2006bv} for static backgrounds, and later generalized in the Hubeny-Rangamani-Takayanagi (HRT) formula\cite{Hubeny:2007xt} for time-dependent backgrounds. The conjectured RT formula has passed various non-trivial checks\cite{Headrick:2007km, Headrick:2010zt} known in quantum information theory and has also found numerous intriguing applications; see {\it e.g.}~\cite{Takayanagi:2012kg} for a recent review.

One important property satisfied by the entanglement entropy is known as the strong sub-additivity, henceforth abbreviated as SSA. A quantum system can be described by the density matrix, usually denoted by $\rho$, which is a self-adjoint, positive semi-definite, trace class operator. The entropy of the corresponding system can be described by the von-Neumann formula: $S = - {\rm tr} [\rho \log \rho]$.

Let us imagine a quantum field theory on a Lorentzian manifold and further imagine a Cauchy surface that divides the entire system in two sub-systems: $A$ and $A^c$ respectively.\footnote{Note that there can be multiple Cauchy surfaces resulting in the same partitioning of the Hilbert space. Thus the Hilbert subspace is specified by the (future) Cauchy horizon rather than the Cauchy surface itself\cite{Casini:2003ix}.} We can now define a ``reduced" density matrix for the sub-system $A$ by tracing over $A^c$: $\rho_A = {\rm tr}_{A^c} \left[ \rho\right]$, and subsequently define a von-Neumann entropy: $S_A = - {\rm tr} \left[ \rho_A \log \rho_A \right]$ as the entanglement entropy. We can now imagine partitioning the Hilbert space by more than one Cauchy surfaces. Specifically, if we have three sub-systems $A_1$, $A_2$ and $A_3$, then SSA is defined as
\begin{eqnarray}
&& S_{A_1 \cup A_2} + S_{A_2 \cup A_3} - S_{A_2} - S_{A_1 \cup A_2 \cup A_3}  \ge 0 \ , \label{ssa1} \\
&& S_{A_1 \cup A_2} + S_{A_2 \cup A_3} - S_{A_1} - S_{A_3} \ge 0 \ . \label{ssa2}
\end{eqnarray}
This property was originally proved in \cite{Araki:1970ba, Lieb:1973cp}, for a recent expository account see {\it e.g.}~\cite{Nielsen:2004}. This inequality, that stands as a cornerstone of quantum information theory, can be viewed as a crucial ingredient in characterizing the von-Neumann entropy\cite{Aczel:1974, Ochs:1978}.

In AdS/CFT correspondence, in a $(d+1)$-dimensional bulk theory the RT formula to compute entanglement entropy of a region $A$ is given by
\begin{eqnarray} \label{hrt}
S_A = \frac{1}{4 G_N^{(d+1)}} {\rm min} \left[ {\rm Area} \left(\gamma_A \right)\right]  ,
\end{eqnarray}
where $G_N^{(d+1)}$ is the bulk Newton's constant, $\gamma_A$ denotes the $(d-1)$-dimensional minimal area surface that satisfies $\partial \gamma_A = \partial A$. For backgrounds with time dependence this proposal is generalized to, {\it via} the HRT formula, considering extremal surfaces rather than a minimal one.\footnote{In case there are more than one extremal surfaces, one chooses the surface with the minimum area.} In \cite{Headrick:2007km}, a simple geometric proof was constructed showing that the RT formula obeys the SSA condition, further substantiating the validity of the RT formula itself.

On the other hand, time-dependent backgrounds do provide a more non-trivial check of the SSA condition. The prototypical example is the so called AdS-Vaidya background, which describes the collapse of a null dust and the formation of a black hole in an asymptotically AdS-space. In the dual field theory this corresponds to a ``global quench" process\footnote{Strictly speaking  we are not really  studying  a  quench process --where a sudden change in a parameter of the Hamiltonian is followed by a unitary evolution.} corresponding to the time evolution from a ``low temperature" state to a thermalized state at a higher temperature. In such a time-dependent background, it was shown in \cite{Allais:2011ys, Callan:2012ip} that violation of SSA is strongly tied to the violation of null energy condition (NEC) from the bulk point of view.\footnote{A violation of the NEC violates the bound in (\ref{ssa1}), whereas (\ref{ssa2}) remains satisfied. Thus, from a holographic perspective, there is a clear distinction between the inequalities in (\ref{ssa1}) and (\ref{ssa2}).} For this example, the background is characterized by a time-dependent mass function and the NEC imposes a condition on this function.

Before proceeding further, let us mull over a curious aside. Within AdS/CFT, the importance of the NEC has been realized elsewhere: in constructing a monotonically decreasing central charge function along an RG-flow\cite{Myers:2010xs,Sinha:2010pm,Myers:2010tj} for a CFT living in arbitrary dimensions. For a $(1+1)$-dim CFT, it can be shown that the SSA condition indeed implies the Zamolodchikov c-theorem\cite{Casini:2004bw}; for more recent developments in higher dimensions see {\it e.g.}~\cite{Casini:2012ei}. Thus, fundamental ``inequalities" in a large $N$ gauge theory, {\it e.g.}~a monotonically decreasing central charge along an RG-flow or the SSA condition, seem to be stemming from the NEC condition in the bulk description.

In this article, we intend to sharpen the connection of the SSA condition with the NEC condition by studying the formation of a charged black hole in AdS-space. In the dual field theory, this will correspond to a global thermalization process in the presence of a chemical potential\cite{Caceres:2012em, Galante:2012pv,Albash:2010mv}. The corresponding background is a Reissner-Nordstr\"{o}m-Vaidya background in AdS-space, henceforth abbreviated as AdS-RN-Vaidya background. This background is characterized by time-dependent mass and charge functions and the corresponding null energy condition has subtle implications. For a given mass and a given charge function, the null energy condition yields a critical surface, denoted by $z_c$, that separates the entire spacetime in two regimes: for $z< z_c$, the NEC is satisfied and for $z>z_c$ it is violated.\footnote{We are working in a coordinate where the boundary of the spacetime is located at $z \to 0$.} Therefore, a violation of the NEC depends on whether the regime $z>z_c$ is accessible to an asymptotic observer.

It was argued in \cite{Israel:1966rt, Ori:1991} that for arbitrary\footnote{The functions are not completely arbitrary; namely, we need to still impose the same condition on the mass function that the NEC imposes in the uncharged AdS-Vaidya case.} mass and charge functions, time-like and null geodesics cannot penetrate the critical surface and hence from a gravitational perspective the NEC is protected by having a no-go constraint on these geodesics. However, in AdS/CFT correspondence, space-like geodesics\footnote{Through this work, when we mention geodesics, we actually mean extremal area surfaces.} are also relevant since they carry the information about non-local operators in the dual field theory such as a $2$-point function, Wilson loop or entanglement entropy. In this article, we will discuss various examples where space-like geodesics can or cannot penetrate this critical surface depending on the choices for the mass and the charge functions. In the dual field theory, this penetration is perceived as a violation of the SSA condition. Thus, we cannot conspire to have a large $N$ gauge theory with certain choices for the mass and the charge functions: entanglement entropy {\it knows it all}. However, we will merely discuss generic and instructive examples rather than attempting for a general characterization of these functions.

This article is divided in the following parts: We begin with a short review about the SSA and the NEC condition in the AdS-Vaidya background in section 2. Then, we discuss in details the AdS-RN-Vaidya background in section 3. We also discuss generic examples relating the physics of the NEC condition with the SSA condition based on our numerical explorations. We provide examples for asymptotically AdS$_4$ and AdS$_5$-backgrounds. Finally, we conclude in section 4.

\section{A brief review}

We begin by briefly reviewing the results that are already known in the literature, specially in \cite{Allais:2011ys, Callan:2012ip}.

\subsection{Strong subadditivity, concavity and monotone-increasing}

Let us begin by demonstrating the relation of concavity and monotone-increasing with the SSA conditions. We will follow closely the discussion in \cite{Callan:2012ip}. Let us consider three adjacent single intervals $A_1$, $A_2$ and $A_3$, whose lengths are denoted by $a_1$, $a_2$ and $a_3$. By symmetry of the construction, $S(A_i) = S(a_i)$. Now, let us assume that $S$ is a concave function. By definition
\begin{eqnarray}
S\left(y x_1 + (1-y) x_2 \right) \ge y S(x_1) + (1-y) S(x_2) \ , \quad 0 < y < 1 \ .
\end{eqnarray}
Now, let us choose $y = a_3/(a_1 + a_3)$
\begin{eqnarray}
&& x_1 = a_2 \ , \quad x_2 =\sum_i a_i \ , \quad \implies S(a_1 + a_2 ) \ge y s(a_2) + (1-y) S\left(\sum_i a_i \right) \ , \label{rel1}\\
&& x_2 = a_2 \ , \quad x_1 =\sum_i a_i \quad \implies S(a_2 + a_3 ) \ge (1-y) s(a_2) + y S\left(\sum_i a_i \right) \ . \label{rel2}
\end{eqnarray}
Adding (\ref{rel1}) and (\ref{rel2}) we get (\ref{ssa1}).

On the other hand, the condition of monotone-increasing yields
\begin{eqnarray}
S(a_1+a_2) \ge S(a_1) \ , \quad S(a_2 + a_3) \ge S(a_3) \ ,
\end{eqnarray}
adding which we get (\ref{ssa2}). Thus the SSA conditions are equivalent to concavity and monotone-increasing conditions on entanglement entropy.

\subsubsection{AdS-Vaidya background and entanglement entropy}

 The AdS-Vaidya metric in a $(d+1)$ spacetime is given by
 \be
 ds^2\,=\,\frac{L^2}{z^2}\left[ - f(z, v) dv^2 -2 dz dv + d\vec{x}^2  \right] \ ,  \quad f = 1-m(v)z^d \ , \label{vaidya}
 \ee
 which describes the formation of a black holes as a shell of null dust collapses. Here, $m(v)$ is a function that interpolates between empty AdS and an AdS-Schwarzschild background as a function of $v$. Also, $L$ is the radius of curvature, $\vec{x}$ is a $(d-1)$-dimensional vector. We have expressed the above background in the Eddington-Finkelstein coordinates, where the coordinate $v$ is defined as
\begin{eqnarray}
dv = dt - \frac{dz}{f(z,v)} ,
\end{eqnarray}
and $t$ denotes the boundary time. In this coordinate system, the boundary is located at $z\to 0$.

The energy-momentum tensor that sources this metric has only one non-vanishing component:
 \be
T_{vv} = \frac{d-1}{2}z^{d-1}\partial_{v} m(v) \ .
\ee
The null energy condition imposes the following constraint on $m(v)$:
\be
T_{\mu \nu} n^\mu n^\nu \geq 0 \quad \implies \quad  \partial_v m(v) \geq 0 \ , \label{NEC}
\ee
where $n^\mu$ is a null vector.

We now use the HRT formula (\ref{hrt}) to compute the entanglement entropy for a spatial region $A$. Let's assume $A$ to be a $(d-1)$-dimensional ``rectangle" in the boundary such that $x^1 \in (-\ell/2,\ell/2)$ and $x^2,...,x^{d-1}\in (0, \ell_{\perp})$ at some fixed boundary time $t_b$. The HRT prescription establishes  that $S_A$ is proportional to the area of the extremal surface $\gamma_A$, parametrized by $v(x)$ and $z(x)$, and whose boundary coincides with the boundary of $A$ at $z=0$.

Thus, we extremize the area
\be
{\rm Area}(\gamma_A) = L^{d-1} \mathcal{V} \int_{-\ell/2}^{\ell/2} dx \frac{1}{z^{d-1}} \sqrt{1- [1-m(v)z^d](v')^2-2 v'z' }, \qquad '\equiv d/dx , \label{area1}
\ee
where $\mathcal{V}\equiv \ell_{\perp}^{d-2} $. We also impose the boundary conditions
\be
v(-\ell/2) = v(\ell/2) = t_b\ , \qquad z(-\ell/2)=z(\ell/2) = 0 \ . \label{bc1}
\ee
The two equations of motion that follow from (\ref{area1}) are
\begin{align}
0 =& \,[1-m(v)z^d]v''+z''-\frac{\partial_v m(v)}{2} z^d (v')^2-d m(v) z^{d-1}z'\,v' \ , \label{eom1}\\
0 =& \,z\, v'' -\frac{d-2}{2}m(v) z^d (v')^2 +(d-1)[(v')^2+dv'z'-1] \ , \label{eom2}
\end{align}
and the momentum conservation corresponding to the cyclic coordinate $x$ results in the equation
\be
1-[1-m(z)z^d](v')^2 -2v'z'=\left( \frac{z_*}{z}\right)^{2(d-1)}  , \qquad z_* \equiv z(0) \ . \label{conse-momentum}
\ee

It can be shown that only two of the above three equations are independent. Thus, we can solve one of the equations of motion together with the conservation equation by imposing appropriate boundary conditions. It is more practical to use the infra-red boundary conditions $z(0)=z_*$, $v(0)=v_*$\footnote{Note that these boundary conditions are guaranteed because of the symmetry of our construction under $x^1 \to - x^1$. The smoothness of the surface at $z_*$ also imposes: $z'(0)=0$ and $v'(0)=0$. Thus, we have sufficient number of boundary conditions altogether.} to solve the equations and, then, read off the values of $t_b$ and $l$ from (\ref{bc1}). Once we solve the system, the extremal area can be computed by simplifyng (\ref{area1}) using (\ref{conse-momentum}):
\be
S(\ell)={\rm Area}(\gamma_A) = 2\,L^{d-1} \mathcal{V} \int_{0}^{\ell/2} dx \frac{z_*^{d-1}}{z^{2(d-1)}} \ . \label{area2}
\ee
This area contains the usual divergent pieces and we will focus on the finite part only.

\subsubsection{Strong subadditivity and the null energy condition}

\begin{figure}[h]
\begin{center}
\unitlength = 1mm

\subfigure[ ]{
\includegraphics[width=60mm]{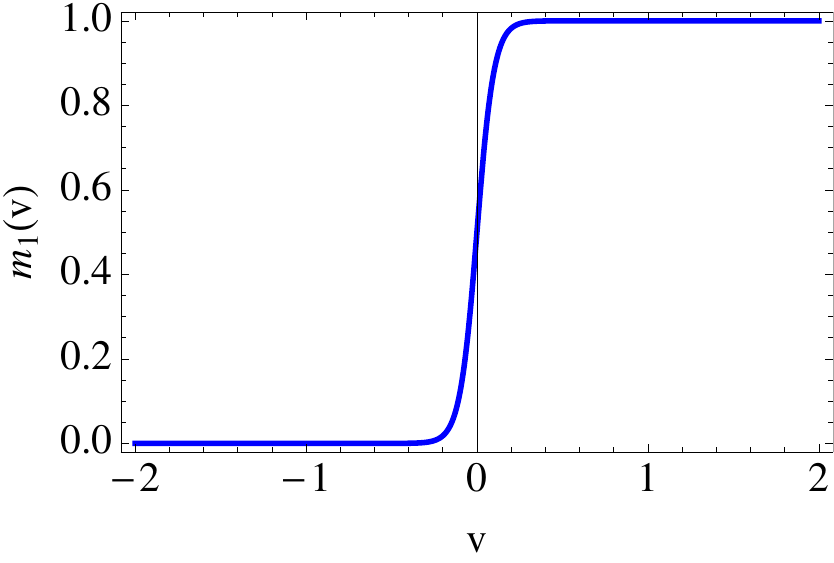}
}
\subfigure[ ]{
\includegraphics[width=60mm]{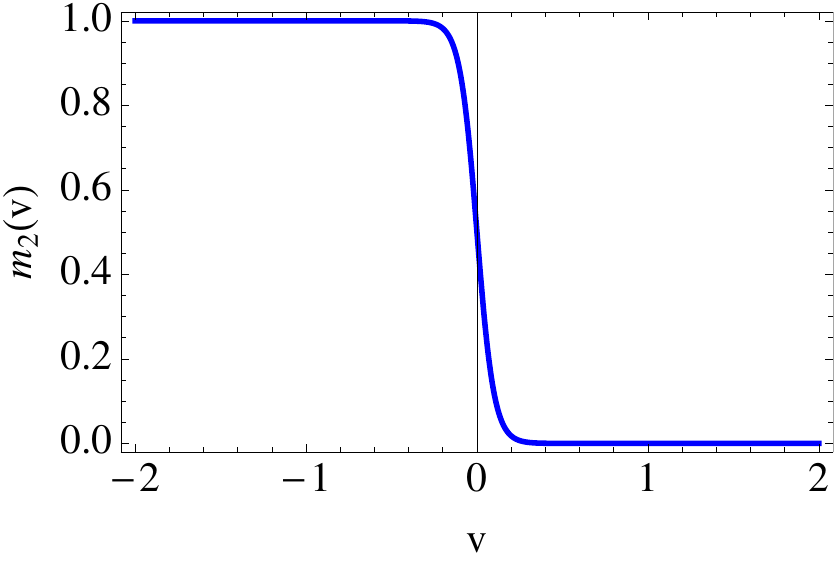}
 }
\caption{\small Examples of two functions $m(v)$ where (a) NEC is obeyed, and (b) NEC is violated.} \label{m(v)functions}
\end{center}
\end{figure}

\begin{figure}[h]
\begin{center}
\unitlength = 1mm

\subfigure[ ]{
\includegraphics[width=67mm]{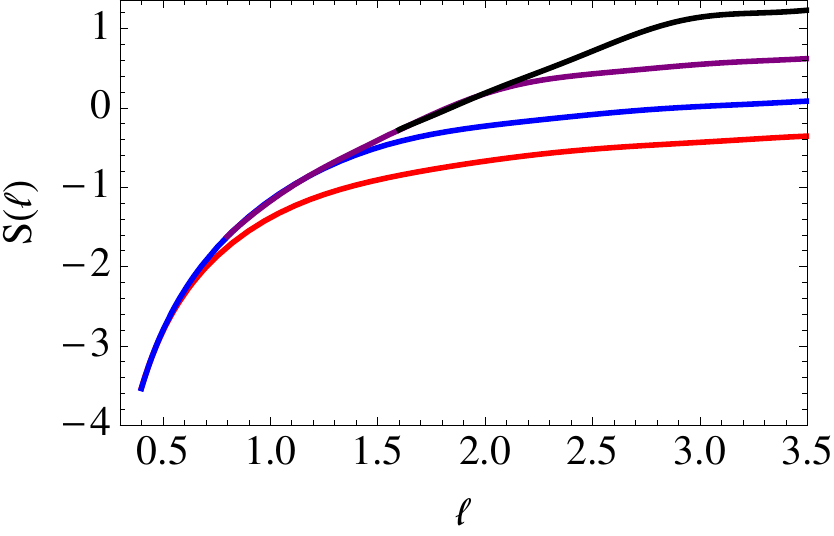}
}
\subfigure[ ]{
\includegraphics[width=67mm]{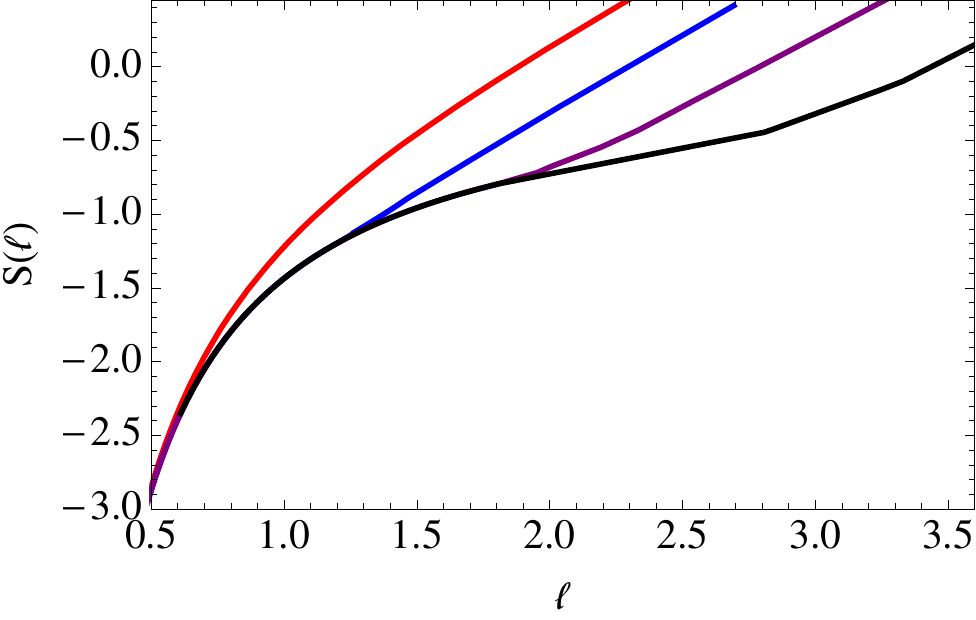}
 }
  \caption{\small Entropy function for the cases where (a) NEC is obeyed, and (b) NEC is violated. The different colors correspond to boundary times $t_b=0.3$ (red), $t_b=1.0$ (blue), $t_b=1.5$ (purple), and $t_b=2.0$ (black). Notice that the curves in (b) are not concave and, thus, SSA is violated.} \label{S(l)functions}
\end{center}
\end{figure}

In order to illustrate the results found in \cite{Allais:2011ys, Callan:2012ip} about the relationship between SSA and NEC, we consider two explicit forms of the function $m(v)$ in  equation (\ref{vaidya}):
\bea
m_1(v) &=& \frac{M}{2}\left( 1+ {\rm tanh} \left(\frac{v}{v_0}\right)\right)  \ , \label{goodm} \\
m_2(v) &= &\frac{M}{2}\left( 1- {\rm tanh }\left(\frac{v}{v_0}\right)\right) \ ,  \label{badm}
\eea
which depicted in Fig (\ref{m(v)functions}). It is, then, easy to realize that the null energy condition in equation (\ref{NEC}) is obeyed by (\ref{goodm}) and violated by (\ref{badm}).
We now specialize in the case $d=3$ and solve equations (\ref{eom2}) and (\ref{conse-momentum}) for both functions $m(v)$ using $M=1.0$, $v_0=0.01$. Then, we plot the entropy function in equation (\ref{area2}) for different values of the boundary time. It is found that for the case obeying NEC, $S(\ell)$ is a monotonically increasing function that is also concave; whereas for the NEC violating function, $S(\ell)$ is still increasing monotonically but it is not a concave function. These results are shown in Figure \ref{S(l)functions}. The fact that $S(\ell)$ is not a concave function leads to conclude that SSA is violated, which establishes a direct connection between SSA and NEC.
%

\section{The AdS-RN-Vaidya background}

We will now delve into discussing how this connection of SSA and NEC manifests itself when there is a non-zero background charge. Here we will flesh out all details in their full glory.

\subsection{The bulk action and the backgrounds}

Our initial goal is to write down a metric which describes the formation of a charged Reissner-Nordstr\"{o}m (RN) black hole in AdS-space. One such time-dependent background that smoothly interpolates between pure AdS to AdS-RN background is given by the so called AdS-RN-Vaidya background. In $(d+1)$-bulk dimensions, this background is given by\footnote{We are considering the case $d>2$. The case of $d=2$ is somewhat special, which we will briefly comment on later.}
\begin{eqnarray}
&& ds^2 = \frac{L^2}{z^2} \left( - f(z,v) dv^2 - 2 dz dv + d\vec{x}^2\right) \ , \quad A_v = q(v) z^{d-2} \ ,  \label{adsrnv} \\
&& f(z,v) = 1 - m(v) z^d + \frac{(d-2) q(v)^2}{(d-1) L^2} z^{2(d-1)} \ , \quad \Lambda = - \frac{d(d-1)}{2 L^2} \ , \label{adsrnvf}
\end{eqnarray}
where $L$ is the radius of curvature, $z$ is the AdS-radial coordinate, $\vec{x}$ is a $(d-1)$-dimensional vector, $A_v$ is the gauge field and $m(v)$ and $q(v)$ are the mass and the charge functions that depend on time. As before, we are working with the Eddington-Finkelstein coordinates. The mass and the charge functions, denoted by $m(v)$ and $q(v)$, are hitherto unconstrained.

The background in (\ref{adsrnv}) can be obtained as a solution to the Einstein-Hilbert-Maxwell action with a negative cosmological constant coupled to an external source
\begin{eqnarray}
&& S = S_0 + \kappa \, S_{\rm ext} \ , \\
&& S_0 = \frac{1}{8\pi G_N^{(d+1)}} \left[ \frac{1}{2} \int d^{d+1}x \sqrt{-g} \left(R - 2 \Lambda \right) - \frac{1}{4} \int d^{d+1}x \sqrt{-g} F_{\mu\nu}F^{\mu\nu} \right ] \ ,
\end{eqnarray}
where $S_0$ denotes the Einstein-Hilbert-Maxwell term, $S_{\rm ext}$ denotes the external source and $\kappa$ is some coupling.

The equations of motion resulting from this action are given by
\begin{eqnarray}
&& R_{\mu\nu} - \frac{1}{2} \left(R - 2 \Lambda \right) g_{\mu\nu} - g^{\alpha\rho} F_{\rho\mu} F_{\alpha\nu} + \frac{1}{4} g_{\mu\nu} F_{\alpha\beta}F^{\alpha\beta} = 2 \left(8 \pi G_N^{(d+1)} \kappa \right)  T_{\mu\nu}^{\rm ext} \ , \\
&& \partial_{\rho} \left( \sqrt{-g} g^{\mu\rho} g^{\nu\sigma} F_{\mu\nu}\right)  = \left(8 \pi G_N^{(d+1)} \kappa \right) J_{\rm ext}^\sigma \ ,
\end{eqnarray}
where the external energy-momentum tensor $T_{\mu\nu}^{\rm ext}$ and the external current $J_{\mu}^{\rm ext}$ are contained within the action $S_{\rm ext}$. More precisely, in order for the equations of motion to be satisfied, we have
\begin{eqnarray}
&& 2\kappa T_{\mu\nu}^{\rm ext} = \left(\frac{d-1}{2} z^{d-1}\frac{dm}{dv} - \frac{d-2}{L^2} z^{2d-3} q(v) \frac{dq}{dv} \right) \delta_{\mu v} \delta_{\nu v} \ , \label{tmunu}\\
&& \kappa J_{\rm ext}^\mu = (d-2) L^{d-3} \frac{dq}{dv} \delta^{\mu z} \ .
\end{eqnarray}

In the special case for $d=2$, we get
\begin{eqnarray}
&& ds^2 = \frac{L^2}{z^2} \left(- f(z,v) dv^2 - 2 dv dz + dx^2 \right) \ , \quad A_v = q(v) \log z \ , \label{adsrnv2} \\
&& f(z,v) = 1 - m(v) z^2 + \frac{q(v)^2}{L^2} z^2 \log z \ , \quad \Lambda = - \frac{1}{L^2} \ .
\end{eqnarray}
The above background is sourced by the following energy-momentum tensor and vector current
\begin{eqnarray} \label{tj2}
2 \kappa T_{\mu\nu}^{\rm ext} = \frac{z}{2} \left(\frac{dm}{dv} - \frac{2}{L^2} q(v) \frac{dq}{dv} \log z\right) \delta_{\mu v}\delta_{\nu v} \ , \quad \kappa J_{\rm ext}^{\mu} = \frac{1}{L} \frac{dq}{dv} \delta^{\mu z} \ .
\end{eqnarray}
There is a subtlety in the identification of the source and the VEV in this case and the chemical potential should be identified with the sub-leading term rather than the leading term of the gauge field as one approaches the boundary. For a detailed discussion on this, see \cite{Jensen:2010em}. However, we will not discuss this case.

The energy-momentum tensor presented in equations (\ref{tmunu}) and (\ref{tj2}) corresponds to the energy-momentum tensor of a charged null dust. The easiest way to understand this is to note that
\begin{eqnarray}
T_{\mu\nu}^{\rm ext} \sim k_\mu k_\nu \ , \quad {\rm with} \quad k^2 = 0 \ ,
\end{eqnarray}
where we have chosen the vector $k_\mu = \delta_{\mu v}$, which is a lightlike vector.

Now, NEC --- which a reasonable energy-momentum tensor should obey --- is given by the following inequality: $T_{\mu\nu}^{\rm ext} n^\mu n^\nu \ge 0$, where $n^\mu$ is lightlike, {\it i.e.}~$n^\mu n_\mu =0$. There are two solutions to the null normal equation $n^\mu n_\mu =0$. Without any loss of generality we can write them as
\begin{eqnarray}
n_{(1)}^\mu = \left(0 , 1, \vec{0} \right) \ , \quad n_{(2)}^\mu = \left( 1 , - \frac{1}{2} f , \vec{0} \right) \ ,
\end{eqnarray}
where $\vec{0}$ denotes the components along the $\vec{x}$-directions, which we have chosen to set to zero. The null vector $n_{(1)}^\mu$ imposes a trivial constraint, and the null vector $n_{(2)}^\mu$ imposes
\begin{eqnarray}
&& \frac{d-1}{2} z^{d-1}\frac{dm}{dv} - \frac{d-2}{L^2} z^{2d-3} q(v) \frac{dq}{dv} \ge 0 \ , \quad {\rm for} \quad d>2 \ ,  \label{necgen} \\
&& \frac{dm}{dv} - \frac{2}{L^2} q(v) \frac{dq}{dv} \log z \ge 0 \ , \quad {\rm for} \quad d=2 \ . \label{nec2}
\end{eqnarray}
Clearly, NEC is obeyed for all $z > z_c$, where $z_c$ denotes the radial position beyond which the null energy condition is violated. This critical surface is given by
\begin{eqnarray}
&& z_c^{d-2} = \frac{d-1}{d-2} \frac{L^2}{2} \frac{m'}{q q'} \ , \quad {\rm for} \quad d>2 \ , \label{zc} \\
&& \log z_c = \frac{L^2}{2} \frac{m'}{q q'} \ , \quad {\rm for} \quad d=2  \ .
\end{eqnarray}
Here $' \equiv d/dv$.

A few comments are in order. Note that for the neutral case, we have $q(v) = 0 = q'(v)$ identically. In that case, the critical surface does not exist. In turn, the null energy condition then imposes a condition on the mass function: $m'(v) \ge 0$. In \cite{Allais:2011ys, Callan:2012ip}, this condition was related to the strong sub-additivity property of entanglement entropy, which we reviewed in the previous section. In the absence of charge, however, the null energy condition seems to be correlated with other simple observations as well. We will discuss these momentarily.

Before proceeding further, let us introduce the apparent horizon for the backgrounds in (\ref{adsrnv}) and (\ref{adsrnv2}) following the notations in \cite{Figueras:2009iu}. The apparent horizon is given by the null hypersurface which has vanishing expansion of the outgoing null geodesics. For these backgrounds, the tangent vectors to the ingoing and the outgoing null geodesics are
\begin{eqnarray}
\ell_- = - \partial_z \ , \quad \ell _+ = - \frac{z^2}{L^2} \partial_v + \frac{z^2}{2 L^2} f \partial_z \
\end{eqnarray}
such that we satisfy
\begin{eqnarray}
\ell_- \cdot \ell_- = 0 = \ell_+ \cdot \ell_+ \ , \quad \ell_- \cdot \ell_+ = - 1 \ .
\end{eqnarray}
The codimension $2$ spacelike hypersurface, which is orthogonal to the above null geodesics, has an area: $\Sigma = (L/z)^{d-1}$. The expansion parameters associated with this hypersurface are
\begin{eqnarray}
\theta_{\pm} = \cL_{\pm} \log \Sigma = \ell_{\pm}^\mu \partial_\mu \left( \log \Sigma \right) \ ,
\end{eqnarray}
where $\cL_{\pm}$ denotes the Lie derivatives along the null directions $\ell_{\pm}$. The location of the apparent horizon is then obtained by solving $\Theta = 0$, where $\Theta = \theta_- \theta_+$. In this particular case, the equation $\Theta =0$ implies $f(z,v) = 0$.

In the absence of any charge, we write down a general treatment including the $d=2$ case. The equation for determining the apparent horizon then yields
\begin{eqnarray}
1 - m(v) z^d = 0 \quad \implies \quad z_{\rm ah} = m(v)^{-1/d} \ .
\end{eqnarray}
Here $z_{\rm ah}$ denotes the apparent horizon. Note that in the future infinity, {\it i.e.}~$v \to\infty$, the apparent horizon coincides with the actual event-horizon.

Note that, during the time-evolution, a global event-horizon exists in the background. This is generated by null geodesics in the background and is the boundary of a causal set. Since the background in (\ref{adsrnv}) has $(d-1)$ Killing vectors $(\partial/\partial x)^a$, the location of the event-horizon is given by a curve $z(v)$. The null geodesic equation in the background (\ref{adsrnv}) is given by
\begin{eqnarray}
\frac{dz_{\rm eh}(v)}{dv} = - \frac{1}{2} f \left(z_{\rm eh}(v),v \right) \ ,
\end{eqnarray}
where $z_{\rm eh}$ denotes the location of the event-horizon. In the limit $v \to + \infty$, we have $z_{\rm ah} = z_{\rm eh}$; however, this is not true in the $v\to - \infty$ limit, {\it i.e.}~the event-horizon lies above the apparent horizon.

It was argued in \cite{Figueras:2009iu} that, during a time-evolution, it is the apparent horizon rather than the event-horizon that can define a ``thermodynamics". Based on an analogy, we can define a ``temperature function" and an ``entropy  function" in terms of the apparent horizon
\begin{eqnarray}
&& T(v) = - \left. \frac{1}{4\pi} \frac{d}{dz} f(z, v) \right |_{z_{\rm ah}} = \frac{d}{4\pi} m(v)^{1/d} \ , \\
&& S(v) = \left(V_{\mathbb{R}^{d-1}} \right) m(v)^{- (d-1)/d} \ .
\end{eqnarray}
Here $V_{\mathbb{R}^{d-1}}$ denotes the volume of the $\vec{x}$-directions. The temperature function is obtained by computing the surface gravity at the apparent horizon and the entropy function is obtained as the area of the apparent horizon. Clearly, $T(v)$ and $S(v)$ have well-defined thermodynamic meaning in the limit $v\to + \infty$.

Now, taking derivative of these functions with respect to $v$, we get
\begin{eqnarray}
\frac{dT(v)}{dv} \sim \frac{dS(v)}{dv} \sim \frac{dm}{dv} \ge 0 \ , \label{ocond}
\end{eqnarray}
where we have used the constraint coming from the null energy condition in (\ref{NEC}). From the perspective of the boundary theory, if it makes sense to talk about a ``temperature function" or an ``entropy function" as defined above, the null energy condition implies that these must be monotonically increasing. We already remarked that the null energy condition was demonstrated in \cite{Allais:2011ys, Callan:2012ip} to imply the strong sub-additivity property of entanglement entropy. Thus, either all the above observations are physically equivalent or we are unable to separate them for the example we are considering here.

Before proceeding in to the actual computations, let us make some more observations here. If we introduce a charge in the system, the null energy condition no longer imposes a simple constraint on the mass or the charge function. Instead, it seems to give rise to a critical surface $z_c$, above which the energy condition is violated. It was argued in \cite{Ori:1991} that, for such charged backgrounds, null geodesics never penetrate the critical surface $z_c$ and thus the apparent pathology is not relevant. Within the context of AdS/CFT correspondence, spacelike geodesics are also relevant since they contain informations about non-local operators, such as $2$-point function or the entanglement entropy itself. Our goal here will be to analyze further what choices of mass and charge functions actually violate the null energy condition and how this is perceived from the perspective of the boundary theory as a violation of SSA. For now, we will discuss the case when $m' \ge 0$, which will smoothly connect to the known results when the charge vanishes.

\subsection{Tests of strong subadditivity}

We now proceed to study entanglement entropy and the SSA inequality in holographic theories dual to $(d+1)$-dimensional AdS-RN-Vaidya spacetimes. As mentioned before, $d=2$ is somehow special so in this section we will restrict our attention to the cases with $d \ge 3$.

Our starting point is the metric given in (\ref{adsrnv})-(\ref{adsrnvf}). For simplicity, we consider that the region $A$ in the boundary theory is an infinite ``rectangular strip" with $x_1\in(-\ell/2,\ell/2)$ and $x_i\in(-\infty,\infty)$, $\forall\,\, i\neq1$. We will call $x_1\equiv x$ given that this is the only relevant direction and we will denote the transverse directions collectively as $\vec{x}_\perp$. According to the covariant prescription for entanglement entropy, we have to find the surface $\gamma_A$ living in a constant-$t$ slice that extremizes the proper area functional $\mathrm{Area}(\gamma_A)$. This surface is invariant under translation in $\vec{x}_\perp$. Thus, without loss of generality, we can parameterize it with functions $z(x)$, $v(x)$ and boundary conditions
\be
z(\pm\ell/2)=0\quad\text{and}\quad v(\pm\ell/2)=t\,.
\ee
These boundary conditions impose that the boundary of $\gamma_A$ coincides with the boundary of $A$ along the boundary temporal evolution.  The area of this surface is given by the following functional
\be\label{larangian}
S(\ell)=\mathrm{Area}(\gamma_A)=\mathcal{V}L^{d-1}\int_{-\ell/2}^{\ell/2} \frac{dx}{z^{d-1}} \left(1 - f v'^2 - 2 v' z' \right)^{1/2} \,,
\ee
where $\mathcal{V}$ is the volume that result from integrating over the $\vec{x}_\perp$ directions. Since there is no explicit $x$-dependence in the
Lagrangian, the corresponding conservation equation is given by
\be\label{coneq}
1 - f v'^2 - 2 v' z' = \left(\frac{z_*}{z}\right)^{2(d-1)} \ ,
\ee
where $z_*$ is defined through $z(0)=z_*$. The two equations of motion obtained by extremizing the area functional are
\bea
&& z'' + v'' f + z' v' \frac{\partial f}{\partial z} + \frac{1}{2} v'^2 \frac{\partial f}{\partial v} = 0 \ , \label{eq1}\\
&& z v'' + (d-1)\left(v'^2f+2z'v'-1\right)-\frac{1}{2}zv'^2\frac{\partial f}{\partial z} = 0 \ .\label{eq2}
\eea
In particular, note that the first equation is independent of the dimensions.
By taking the derivative of the conservation equation (\ref{coneq}) with respect to $x$ and using one of these two
equations of motion, one obtains the other one. Thus, it is sufficient to consider only (\ref{coneq}) and
{\it e.g.}~(\ref{eq1}) to find $z(x)$ and $v(x)$. We, then, solve numerically these two equations subject to the boundary conditions
\be
z(0)=z_*\,,\quad z'(0)=0\,,\quad v(0)=v_*\,,\quad v'(0)=0 \ .
\ee
In practice, however, we start the integration at some arbitrarily small $x=\epsilon$ to avoid possible numerical issues. Also, due to the symmetry of the problem, it is sufficient to integrate only for positive values of $x$.

So far, $z_*$ and $v_*$ are two free
parameters that generate the numerical solutions for $z(x)$ and $v(x)$. The boundary data $\{\ell, t_b \}$ can be obtained from these numerical solutions through $z(\ell/2)=z_0$ and $v(\ell/2)=t_b$, where $z_0$ is a UV cutoff. This cutoff is needed because, the area functional (\ref{larangian}) is divergent and one needs to regularize. The divergence comes from the fact that the volume of any asymptotically AdS background is infinite and the spatial surface $A$ we are considering reaches the boundary.

The divergence term can be isolated by studying the same problem in AdS$_{d+1}$ in the standard way \cite{Ryu:2006ef}. Parameterizing with functions $x(z)$ and $v(z)$, it is clear that near the boundary $x'(z\to0)=0$, $v'(z\to0)=0$ and therefore
\be
S_{\mathrm{div}}(\ell)=\mathcal{V}L^{d-1}\int_{z\sim z_0}\!\! \frac{dz}{z^{d-1}} =\frac{2\mathcal{V}L^2}{(d-2)z_0^{d-2}}\,.
\ee
Subtracting this divergence, we obtain the finite term of the area which is the main
quantity we are interested in,
\be\label{areareg}
S_{\mathrm{reg}}(\ell)=2\mathcal{V}L^{d-1}\int_{z_0}^{\ell/2} \frac{dx}{z^2} \left(1 - f v'^2 - 2 v' z' \right)^{1/2}-\frac{2\mathcal{V}L^2}{(d-2)z_0^{d-2}}\,.
\ee

In order to find numerical solutions to the system (\ref{coneq})-(\ref{eq1}), we employ a ``shooting'' method. First, we give initial values $z(0)=z_*$, $v(0)=v_*$ and integrate until the the functions hit the boundary. Once we have the profiles $z(x)$ and $v(x)$, we read the boundary values and extract $\ell$ and $t_b$. The numerical implementation of (\ref{areareg}) is straightforward.

In the remaining part of this section, we will consider specific functions for $m(v)$ and $q(v)$ in $d=3,4.$ As advertised in the introduction, we will find that, although there is always a critical surface above which the null energy condition is violated, for some appropriate choices of mass and charge functions the extremal surfaces attached to the boundary never cross into that region and SSA is satisfied.

\subsection{Thin shells and junction conditions}

Before proceeding to specific examples, let us gain some insight into the properties of the critical surface  (\ref{zc}). To this effect we consider the particular case of a thin null shell located at $v=0$ which is the boundary of two static spaces. The conditions to join space-like or time-like hypersurfaces  demand that the two spacetimes  induce the same  metric on the hypersurface  and relate the surface stress energy tensor $S_{ab}$ to the jump of the normal extrinsic curvature $K_{ab}$ across the hypersurface. The issue is more subtle for null shells since the extrinsic curvature no longer carries any transverse geometrical information. For null hypersurfaces, the extrinsic curvature is given by tangential derivatives of the metric and is, thus, necesarily continuos across the shell and cannot be related to the stress energy tensor of the shell $S_{ab}$. A general formalism applicable to  null hypersurfaces was developed in \cite{BarrabesIsrael}. 

 For simplicity, consider $d=3$. The spacetime metric  is  given by 
\begin{eqnarray}
	&& ds^2 = \frac{L^2}{z^2} \left( - f(z,v) dv^2 - 2 dz dv + d\vec{x}^2\right) .  \label{adsrn-thin} 
\end{eqnarray}
\noindent Consider two static backgrounds $\mathcal{M}^-$ and $\mathcal{M}^+$, with mass and charge parameters  $M_i, Q_i$ and  $M_f, Q_f$ respectively. Let $\mathcal{M}^-$ and $\mathcal{M}^+$ be bounded by  hypersurfaces $\Sigma_-$and $\Sigma_+$. We glue the two spaces by identifying $\Sigma_- =\Sigma_+ =\Sigma$. In the present case, we take $\Sigma$ to be the hypersurface $v=0$. We are interested in the case when  $\mathcal{M}^-$ and $\mathcal{M}^+$ are vacuum solutions with,
\begin{align}\label{eq:ftheta}
	 M(v)&=M_i + \Theta(v)(M_f-M_i), \qquad\qquad Q(v)=Q_i +\Theta(v)(Q_f-Q_i) \ ,\\
	 f(z,v)&= 1-M(v)z^3 + \frac{1}{2} Q(v)^2 z^4 \\
	 &= 1 - (M_i + \Theta(v)(M_f-M_i)) z^3 + \frac{1}{2} (Q_i^2+\Theta(v)(Q_f^2-Q_i^2))  z^4 \ ,
\end{align}
where $\Theta(v)$ is the Heaviside step function.  Note that (\ref{eq:ftheta}) allows for the initial and final backgrounds to be   AdS, AdS-Schwarzchild or AdS-RN. The stress energy tensor (\ref{tmunu}) is,
\begin{align}
 2\kappa T_{\mu\nu} &= z^2 \left( \frac{dM}{dv} - z Q(v) \frac{dQ}{dv} \right) \delta_{\mu v} \delta_{\nu v}  \nonumber\\
 &= \delta(v) z^2\Big((M_f-M_i)- z (Q_f-Q_i)(Q_i +\Theta(v)(Q_f-Q_i))\Big )\delta_{\mu v} \delta_{\nu v} \ . \label{tmunudist}
\end{align}
Note that  (\ref{tmunudist}) identically vanishes in $\mathcal{M}^-$ and $\mathcal{M}^+$ and  is non-zero only at  $v=0$. This discontinuity comes with a sound physical interpretation; it is associated with the presence of a thin distribution of matter  at $v=0$. The only non-zero component of the surface stress energy tensor is $T^{z z}$, 
\begin{align}
	2\kappa T^{\mu\nu}_\Sigma &= \delta(v) z^2\Big((M_f-M_i)- z (Q_f-Q_i)(Q_i +\Theta(v)(Q_f-Q_i))\Big ) k^\mu k^\nu \\
	& \equiv \delta(v)
	\sigma(z)\delta^{\mu z} \delta^{\nu z} \ .
\end{align}
Since there is no rest frame for a null shell, we cannot formally identify $\sigma$ as {\it the} surface density. However, $\sigma$ can be used to determine the results of measurements by any specified observer. This involves introducing an arbitrary congruence of timelike geodesics intersecting $\Sigma$ associated with the different families of observers making measurements on the shell. An observer with four-velocity $u^\alpha = dx^\alpha/ d\tau $,  will measure an energy density associated with the shell $T^{\mu\nu}_\Sigma u_\mu u_\nu = \delta(v) \sigma(z) (k^\mu u_\mu)^2 $.  Thus, the arbitrariness of the choice of congruence is limited to an overall factor and the quantity  $\sigma(z)$ is independent of this choice. It is in this sense that we  interpret $\sigma(z) $ as the shell's surface density. 

From (\ref{zc}), we see that for this type of backgrounds  the critical surface is  located at $v=0$ and 
 \begin{equation}\label{eq:zcthin}
	 z_c=\frac{ M_f - M_i}{(Q_f - Q_i)(Q_i +\Theta(0)(Q_f-Q_i))}=\frac{2( M_f - M_i)}{Q_f^2 - Q_i^2}  \ , 
 \end{equation}
 where we have used $\Theta(0)=1/2$ as is conventional in distribution theory. Evaluating $\sigma$ at the critical surface we obtain,
$$\sigma(z_c)=0 \ .$$
Thus, the critical surface is the locus where the shell's surface density becomes zero.

Our main interest is to study extremal spacelike surfaces in the backgrounds described above. From the point of view of the spacelike surface, there is a discontinuity in the $dz/dv$  when the surface crosses the shell. This can easily be seen  from the equations of motion. As in the previous section, consider a rectangular strip in the boundary theory.  Extremizing the area functional (\ref{larangian})  we obtain the  equations of motion  (\ref{eq1}-\ref{eq2})
which for $d=3$ read, 
\begin{align}
	& z v'' + 4 v' z' + (2 f - \frac{z}{2}\frac{df}{dz})v'^2=0 \ , \label{eq:eqgeo1}
\\
& z'' + f v'' + \frac{df}{dz} z' v' + \frac{1}{2} \frac{df}{dv} v'^2=0 \ .\label{eq:eqgeo2}
\end{align}
The $z(x)$ and $v(x)$ coordinates are continuos across the shell while  $f$ and $\frac{df}{dz}$ present a finite jump. From (\ref{eq:eqgeo1}) we see that  $z', v'$ and $ v''$ remain finite. On the other hand,  
\begin{align}
	\frac{d f}{dv}& = z^3 [(M_f-M_i)  -  z (Q_f-Q_i)(Q_i +\Theta(v)(Q_f-Q_i))]\delta(v) \nonumber\\
	&\equiv - \widetilde{ f}(z,v) \delta(v)
\end{align}
diverges at $v=0$. Thus,  the behavior of  (\ref{eq:eqgeo2}) across the shell is,
\begin{align}
	& z''\sim  -\frac{1}{2}  \widetilde{ f}(z,v) \delta(v)
 v'^2
 \end{align}
and  $z'$ has a finite  jump. Indeed,  
 \begin{align}
	 &  ( z')^+ -(z')^- = \int^{0^+}_{0^-} z'' dx =\int^{0^+}_{0^-}  \frac{z''}{v'} dv\sim-\frac{1}{2}\widetilde{ f}(z,0) v'
 \end{align}
 and the jump  of $\frac{dz}{dv}$ across the shell  is,
 $$\Delta \left(\frac{dz}{dv}\right)\equiv \left(\frac{dz}{dv}\right)^+ - \left(\frac{dz}{dv}\right)^-=  -\frac{1}{2}\widetilde{ f}(z,0)  $$
where $+$ and $-$ refer to quantities evaluated in $\mathcal{M}^+$ and $\mathcal{M}^-$ respectively.  Using (\ref{eq:zcthin}) we find that the critical surface corresponds to the surface at which the jump on $\frac{dz}{dv}$ vanishes,

 \begin{equation}
	 \Delta \left(\frac{dz}{dv}\right) \Big|_{z_c}=-\frac{1}{2}\widetilde{ f}(z_c,0) =0 \ .
 \end{equation}
 Intuitively this makes sense;  we know that at $z_c$ the mass of the shell goes to zero, the shell has disappeared and there is no reason for a  jump in $\frac{dz}{dv}$.

\subsection{Examples in $d=3$}

Let us begin our discussion for the $(3+1)$-dimensional bulk theory, where the dual field theory is a $(2+1)$-dimensional conformal field theory in the presence of a chemical potential. Presumably the corresponding UV-completion is given by an $S^7$-reduction of $11$-dimensional supergravity, which leads to an $SO(8)$ gauged supergravity in $(3+1)$-dimensions. Therefore, the boundary theory should correspond to an ABJM-like, {\it i.e.}~Chern-Simons-matter theory in the presence of a chemical potential.\footnote{We should note that while this is a very plausible scenario, we are making an assumption that the Vaidya-type backgrounds can be embedded within gauged supergravity consistently {\it at least} in some well-defined limit.}

\subsubsection{Backgrounds that respect SSA}

We will begin with the examples that do not violate SSA. A good starting point is to use the functions analyzed in \cite{Caceres:2012em} that were used to address scaling properties of the thermalization time with respect to the temperature and the chemical potential of the thermalized state. These are: 
\be
m(v)=\frac{1}{2}\left(1+\tanh\left(\frac{v}{0.01}\right)\right)\qquad\text{and}\qquad q(v)=0.9\, m(v)^{2/3} \ . \label{funct1} 
\ee
\begin{figure*}[h]
$$
\begin{array}{cc}
  \includegraphics[width=7cm]{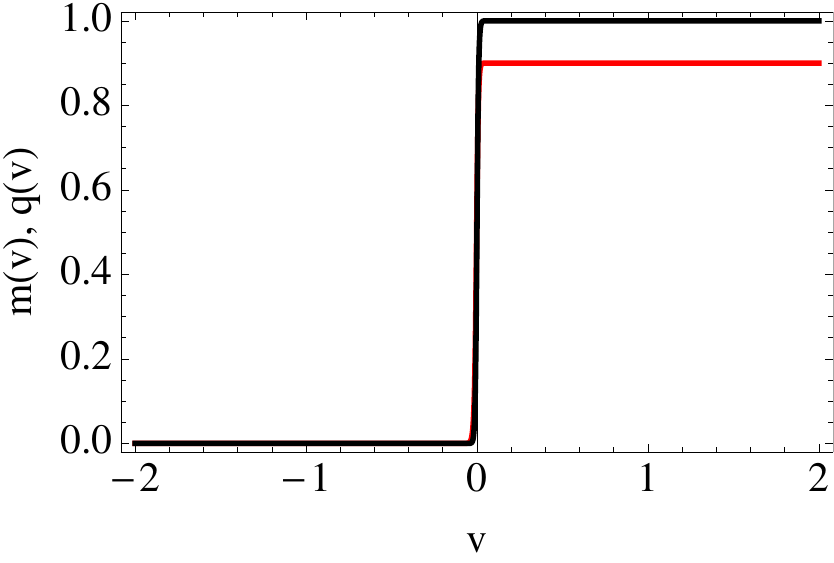} &\quad \includegraphics[width=7cm]{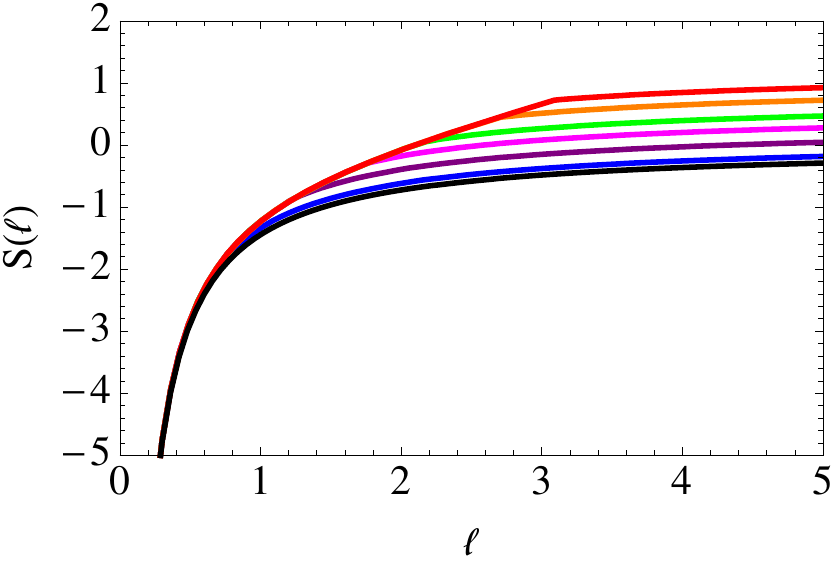}
\end{array}
$$
\caption{Left panel: $m(v)$ (black) and $q(v)$ (red). Right panel: $S(\ell)$, for $t_b=$ 0.01 (black), 0.5 (blue), 1 (purple), 1.5 (magenta), 2 (green), 3 (orange) and 5 (red).} \label{nssa1}
\end{figure*}
It is clear from Figure \ref{nssa1} that there is no change in concavity of the entanglement entropy function as $\ell$ grows for a given boundary time $t_b$.

\begin{figure}[h]
\begin{center}
\unitlength = 1mm

\subfigure[ ]{
\includegraphics[width=60mm]{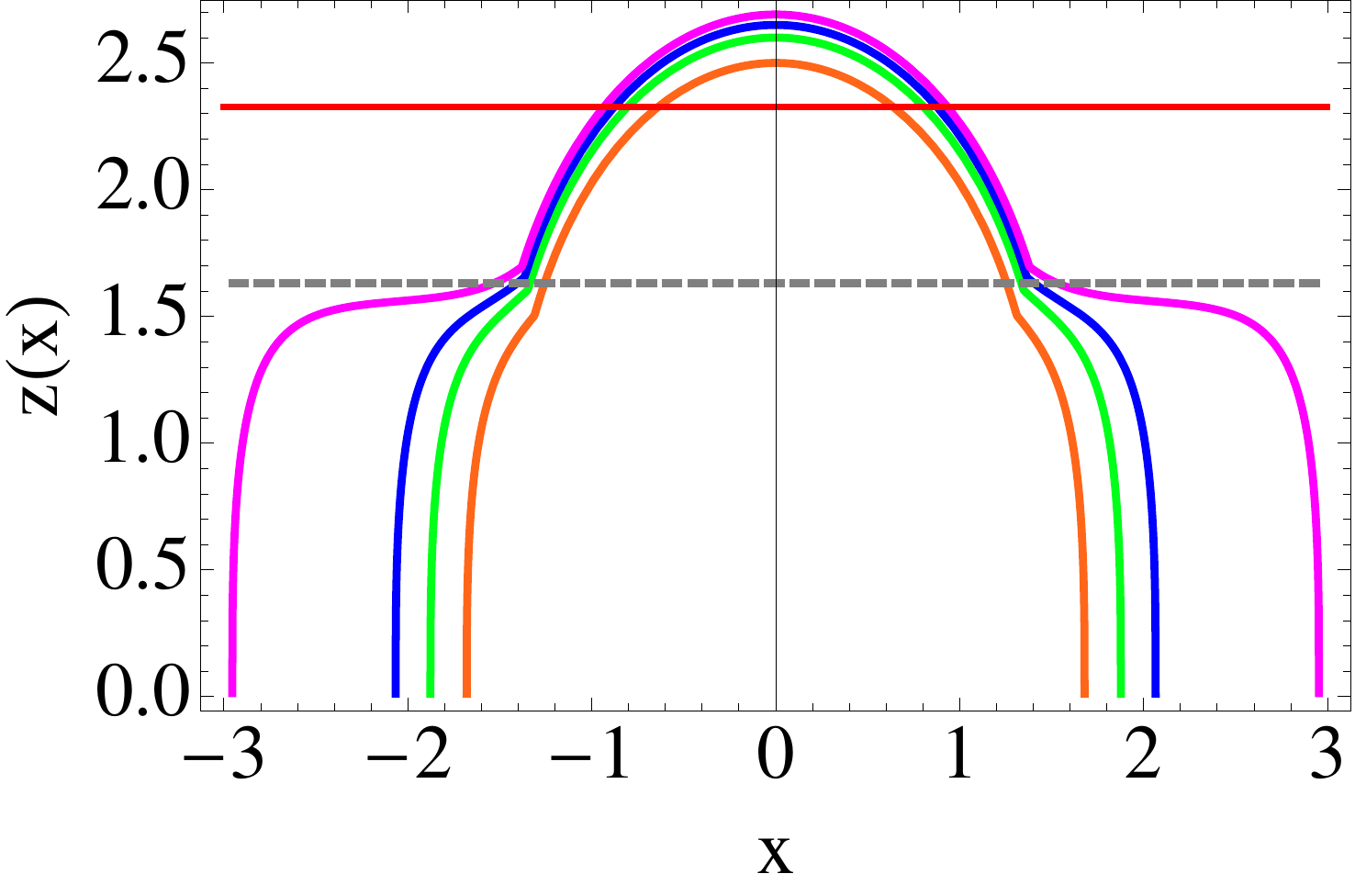}
}
\subfigure[ ]{
  \includegraphics[width=80mm,height=40mm]{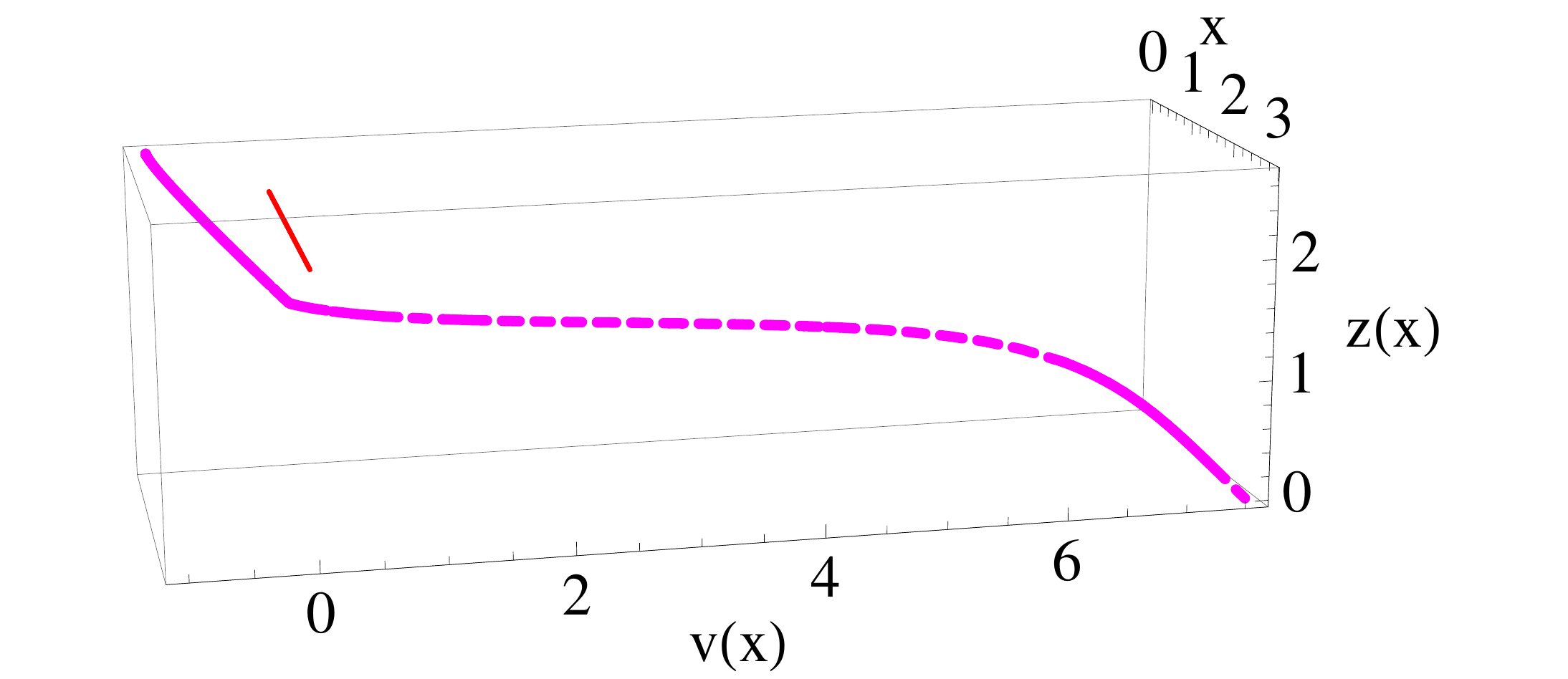}
 }
  \caption{\small Profiles of a family of geodesics when SSA is obeyed. The dashed gray line represents the apparent horizon at $v=0$ and the red line represents the critical surface. This family is parametrized by $\ell$, the length of the entangling region at the boundary. Note that right panel shows that the geodesics do not intersect the critical surface as explained in text.} \label{geod1}
\end{center}
\end{figure}

According to equation (\ref{zc}), there exists a critical surface in the bulk geometry beyond which NEC is violated. It is thus instructive to analyze whether the space-like geodesics, which eventually determine the entanglement entropy, can penetrate this critical surface or not. In Figure \ref{geod1}, we show a representative family (characterized by the length of the entangling region) of geodesics corresponding to function (\ref{funct1}). We also display the critical surface (\ref{zc}) and the location of the apparent horizon.

Before going further we should note that the critical surface, in the thin-shell limit, exists within a tiny\footnote{The width of this region is characterized by the shell thickness parameter $v_0$.} region around $v=0$. Also, we note that the coordinate $v$ evolves along each geodesics independently and at $v=0$ these geodesics cross the shell, which in the thin-shell limit coincides with the apparent horizon. Figure \ref{geod1} therefore compares the location of the critical surface to the location of the geodesics at $v=0$.

There are two key features that stand out from Figure \ref{geod1}: First, the critical surface lies above the apparent horizon\footnote{{\it i.e.} $z_c > z_{ah} $.}; in other words, it is cloaked by the apparent horizon. Second, although the space-like surfaces cross the apparent horizon, they do not probe the {\it forbidden region} beyond the critical surface irrespective of how large $\ell$ becomes. At this point we would like to stress that the above observations seem very robust against a substantial amount of test cases. One might imagine designing a situation where the critical surface comes arbitrarily close to the apparent horizon, thus encouraging the geodesics to cross it. However, we have verified that this does not seem to happen.

Let us now illustrate a couple of more examples where the SSA condition is satisfied. 
\be
m(v)=1+\frac{1}{2}\left(1+\tanh\left(\frac{v}{0.01}\right)\right)\quad\text{and}\qquad q(v)=\frac{0.9}{2}\left(1-\tanh\left(\frac{v}{0.01}\right)\right) \ . \label{exam2}
\ee
\begin{figure*}[h]
$$
\begin{array}{cc}
  \includegraphics[width=7cm]{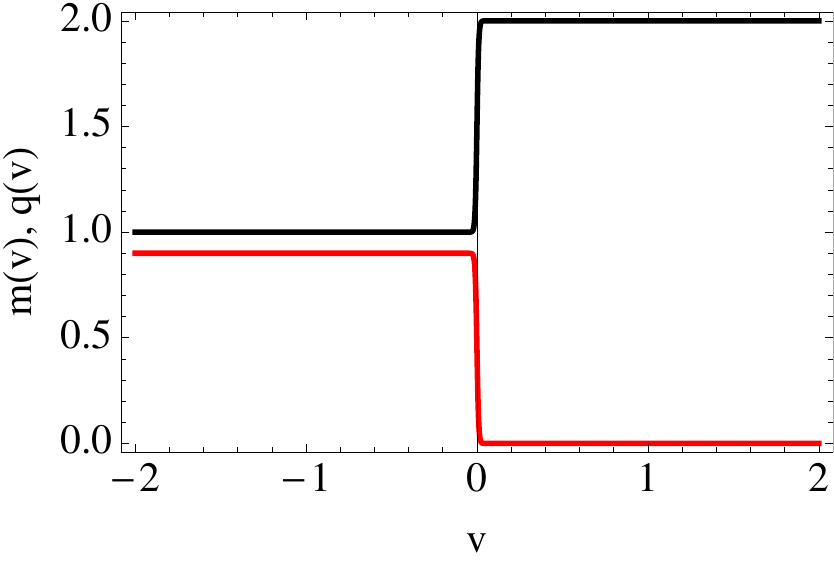} & \includegraphics[width=7cm]{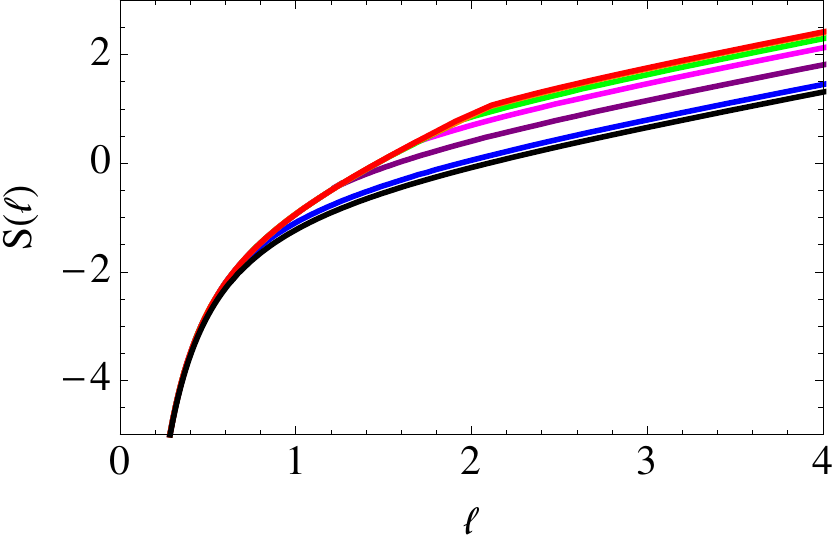}
\end{array}
$$
\caption{Left panel: $m(v)$ (black) and $q(v)$ (red) given in (\ref{exam2}). Right panel: $S(\ell)$, for $t_b=$ 0.01 (black), 0.5 (blue), 1 (purple), 1.5 (magenta), 2 (green), 3 (orange) and 5 (red).} \label{3df1}
\end{figure*}
\be
m(v)=\frac{1}{2}\left(1+\tanh\left(\frac{v}{0.01}\right)\right)\quad\text{and}\qquad q(v)=\frac{0.9}{2}\left(\tanh\left(\frac{v}{0.01}\right)-\tanh\left(\frac{v-1}{0.01}\right)\right) \ . \label{exam3}
\ee
\begin{figure*}[h]
$$
\begin{array}{cc}
  \includegraphics[width=7cm]{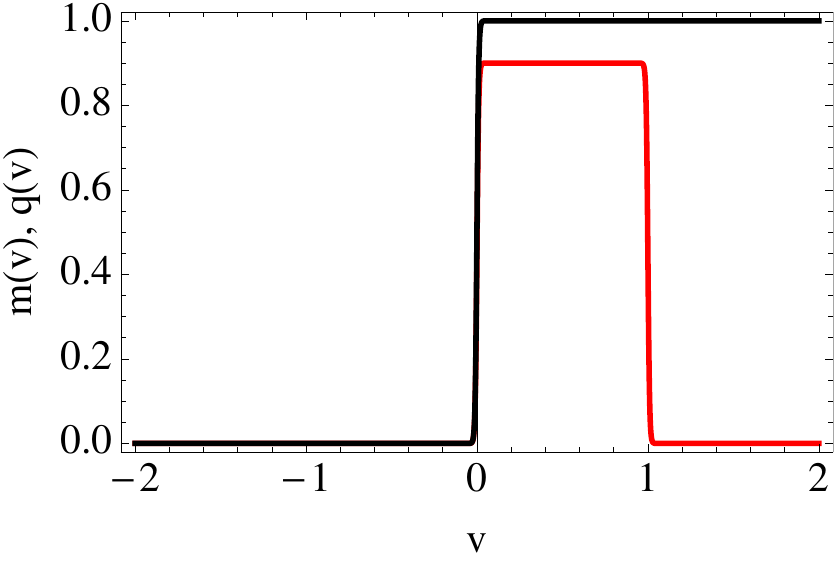} & \includegraphics[width=7cm]{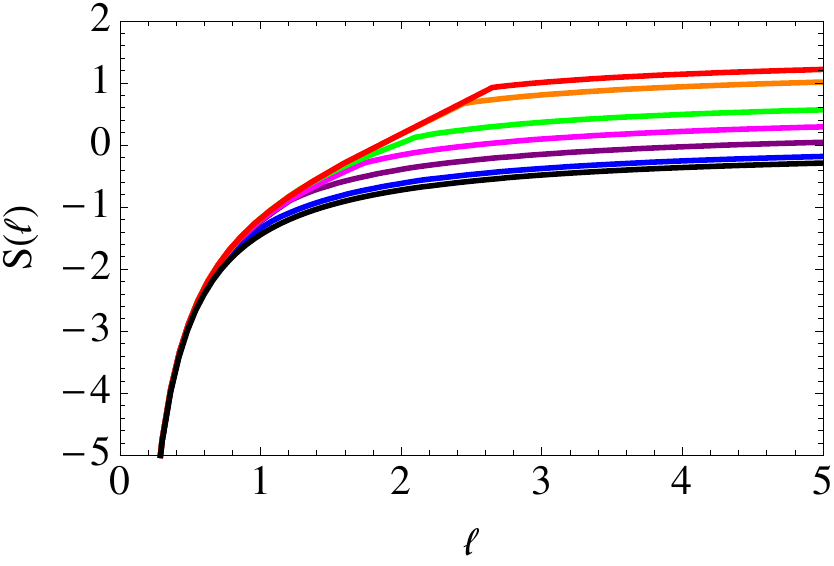}
\end{array}
$$
\caption{Left panel: $m(v)$ (black) and $q(v)$ (red) given in (\ref{exam3}). Right panel: $S(\ell)$, for $t_b=$ 0.01 (black), 0.5 (blue), 1 (purple), 1.5 (magenta), 2 (green), 3 (orange) and 5 (red).} \label{3df2}
\end{figure*}
The choice functions are given in equations (\ref{exam2}) and (\ref{exam3}) and the corresponding figures are shown in Figure \ref{3df1} and \ref{3df2} respectively.

Before concluding this section, let us offer some remarks. In view of (\ref{ocond}), it can be verified that the condition $dT(v)/dv \ge 0$ is not satisfied for all $v$ with the choices made in (\ref{exam3}). Thus the SSA condition is an independent constraint which is not related to the rate of change of surface gravity at the apparent horizon in the dynamical geometry. As far as the corresponding physical processes are concerned, the choices in (\ref{funct1}) represents a situation in the dual field theory, where both temperature and chemical potential are increasing from a ``low value" to a higher non-zero value. On physical grounds, this is perhaps the most ``reasonable" process.

The choices in (\ref{exam2}) takes a low temperature, high chemical potential initial state to a high temperature, low chemical potential final state. Finally, the choices in (\ref{exam3}) are rather exotic, which takes a low temperature, vanishing chemical potential initial state to a high temperature vanishing chemical potential final state; but does not obey $dT(v)/dv \ge 0$, $\forall v$. As far as NEC or SSA is considered, there is nothing preventing these two choices; however, whether they are realizable as solutions of gravity with a reasonable matter field is an issue we will not address here.

\subsubsection{Backgrounds that violate SSA}

\begin{figure*}[h]
$$
\begin{array}{cc}
  \includegraphics[width=7cm]{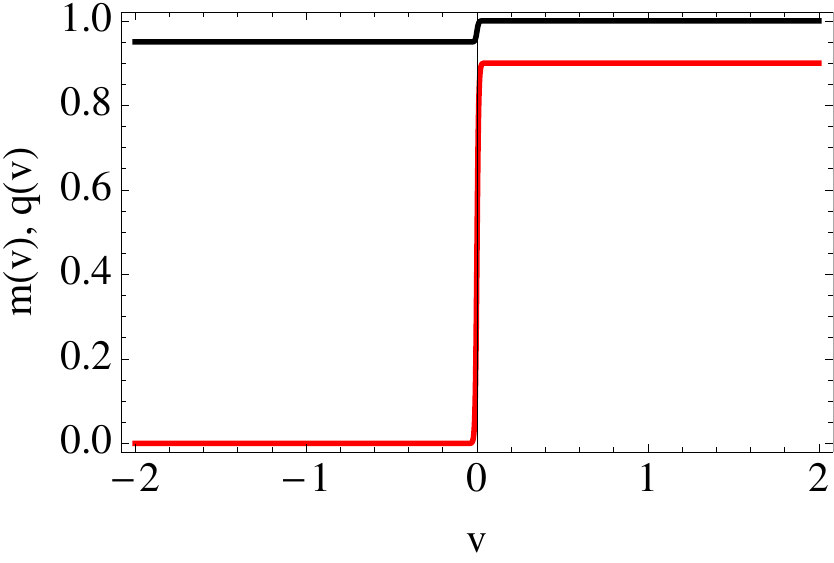} & \includegraphics[width=7cm]{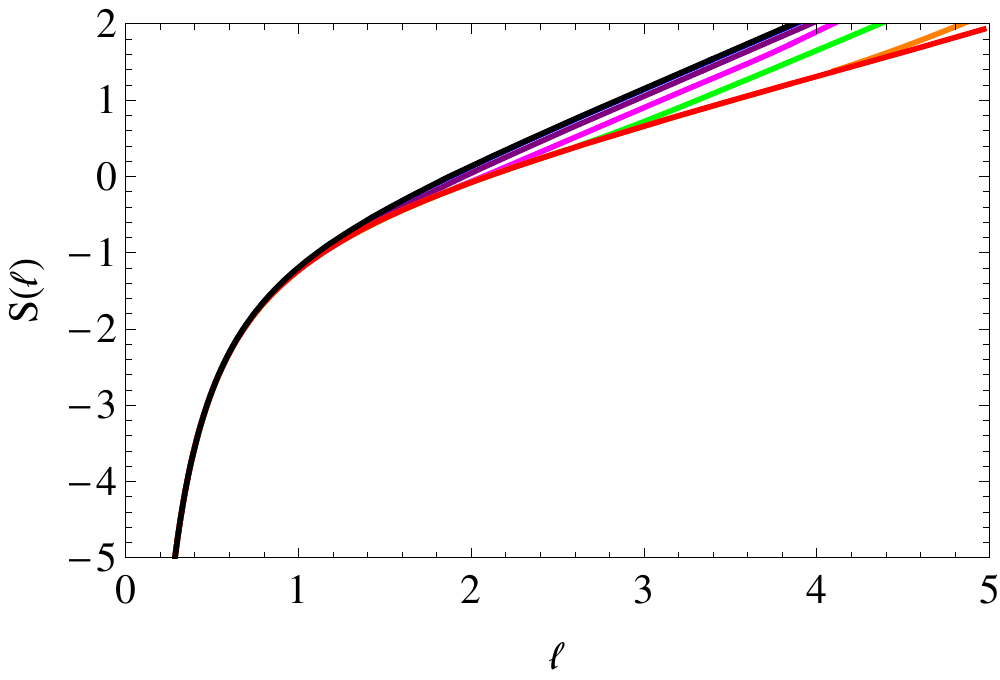}
\end{array}
$$
\caption{Left panel: $m(v)$ (black) and $q(v)$ (red) given in (\ref{exam4}). Right panel: $S(\ell)$, for $t_b=$ 0.01 (black), 0.5 (blue), 1 (purple), 1.5 (magenta), 2 (green), 3 (orange) and 5 (red).} \label{demon2}
\end{figure*}

\begin{figure}[h]
\begin{center}
\unitlength = 1mm
\includegraphics[width=85mm]{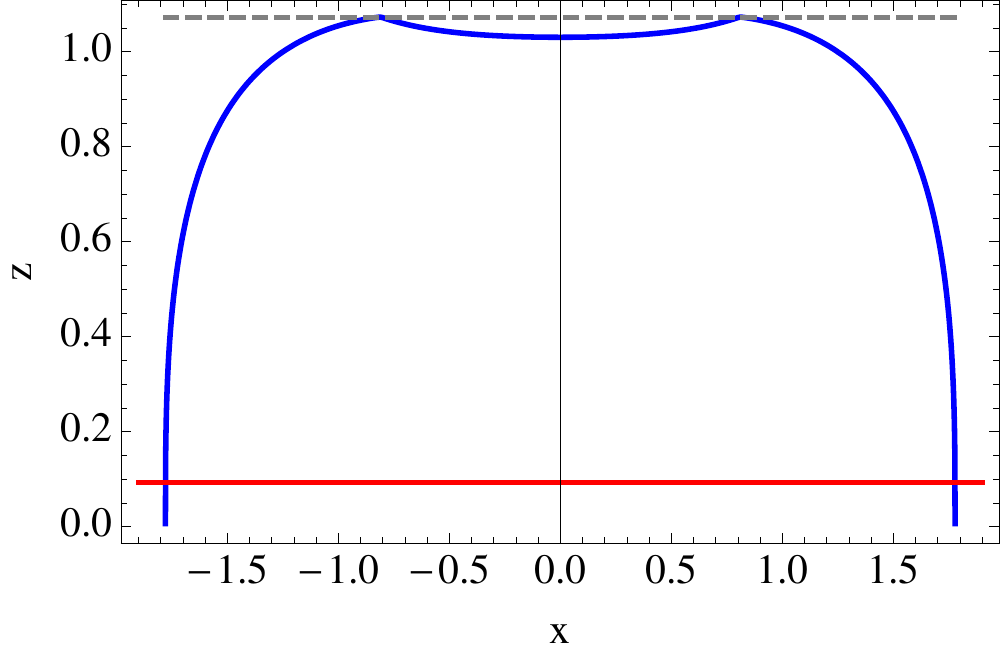}
\caption{\small Profile of geodesic when SSA is violated corresponding to the choice (\ref{exam4}). The dashed line represents the apparent horizon or the position of the shell at $v=0$ and the red line depicts the critical surface at $v=0$. Clearly the geodesic probes the region behind the critical surface at $v=0$.} \label{geod2}
\end{center}
\end{figure}

Now let us illustrate a few examples where SSA is violated. One such choice is:
\be
m(v)=0.95+\frac{0.05}{2}\left(1+\tanh\left(\frac{v}{0.01}\right)\right)\quad\text{and}\qquad q(v)=\frac{0.9}{2}\left(1+\tanh\left(\frac{v}{0.01}\right)\right) \ . \label{exam4}
\ee
The corresponding illustration is shown in Figure \ref{demon2}. The corresponding geodesic is shown in Figure \ref{geod2}. Once again, the critical surface exists only around a small neighbourhood of $v=0$ and the geodesics reach the shell at $v=0$. Note here that both the features alluded to in the previous subsection are gone: First, the critical surface lies outside the apparent horizon at $v=0$; second, the minimal area surface {\it brings news} from the {\it forbidden region} in the bulk by probing the region beyond the critical surface. This is perceived as the violation of SSA condition in the boundary theory. At this point we emphasize that these observations seem rather generic and hence we will not pictorially illustrate a similar behaviour of the geodesics for other representative cases, whenever SSA is violated. 

Let us now take a second example where the violation of SSA is observed: 
\be
m(v)=\frac{1}{2}\left(1+\tanh\left(\frac{v}{0.01}\right)\right)\quad\text{and}\qquad q(v)=\frac{0.9}{2}\left(1+\tanh\left(\frac{v-1}{0.01}\right)\right) \ . \label{exam5}
\ee
\begin{figure*}[h]
$$
\begin{array}{cc}
  \includegraphics[width=7cm]{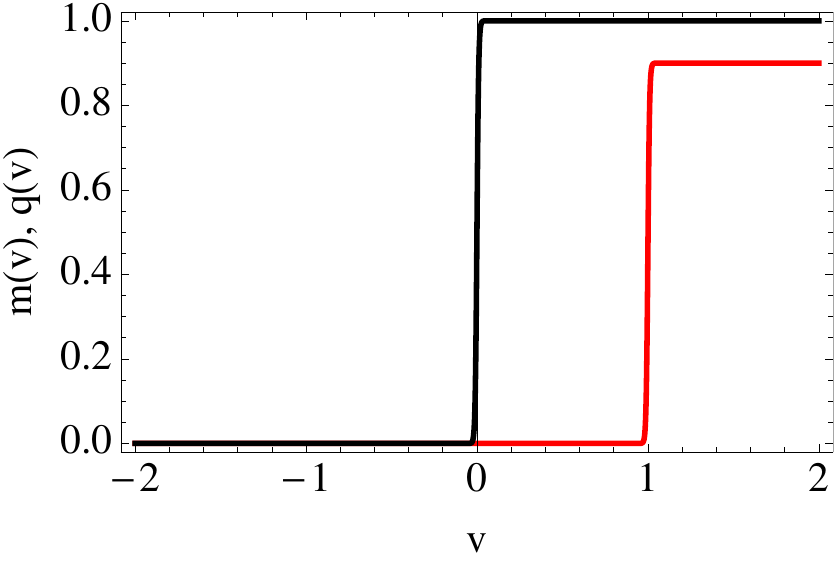} & \includegraphics[width=7cm]{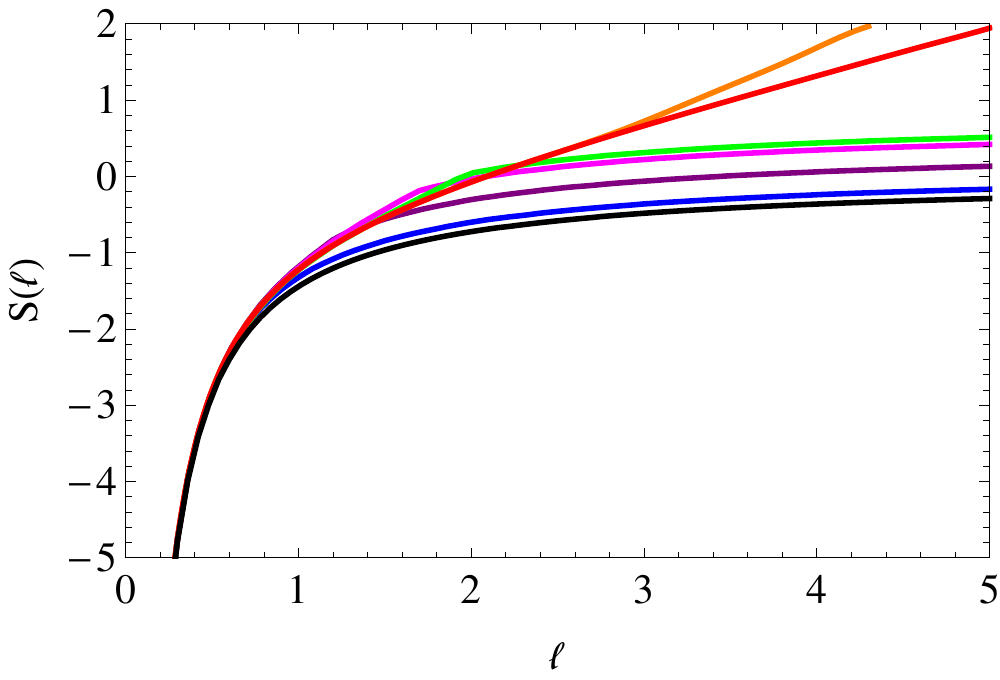}
\end{array}
$$
\caption{Left panel: $m(v)$ (black) and $q(v)$ (red) given in (\ref{exam5}). Right panel: $S(\ell)$, for $t_b=$ 0.01 (black), 0.5 (blue), 1 (purple), 1.5 (magenta), 2 (green), 3 (orange) and 5 (red).} \label{demon3}
\end{figure*}
In this case also the minimal area surface probes the region beyond the critical surface. The corresponding plots showing the violation of SSA are presented in Figure \ref{demon3}.

Once again, a few comments are in order. First, it can be checked that for the choices in (\ref{exam5}) we will get $dT(v)/dv \ge 0$ and $dS(v) /dv \ge 0$ and still a violation of SSA. Thus, the SSA condition is indeed independent of these. Intuitively, the process in (\ref{exam5}) is not very meaningful since it seems to allow for the black hole to accumulate charges keeping it's mass fixed. On the other hand, there is no such {\it a priori} objection to the process in (\ref{exam4}), and still SSA is violated. This indicates that SSA is a rather non-trivial condition on the allowed trajectories a thermalization process might take for a given field theory.

\subsection{Examples in $d=4$}

We will now move up by one dimension and consider a $(3+1)$-dimensional conformal field theory. The dual geometry will correspond to the formation of a charged black hole in five-dimensional AdS-space. In this case, the possible UV-completion will be given by an $S^5$-reduction (or a reduction on a Sasaki-Einstein five-manifold) of type IIB supergravity truncated to the ${\cal N}=2$ sector with an $U(1)^3$ symmetry (or at least one $U(1)$). Thus the dual field theory is presumably a cousin of the prototype ${\cal N}=4$ super Yang-Mills (SYM) theory in the presence of a chemical potential.\footnote{Once again we note that this claim needs to be rigorously demonstrated, which we will not attempt here.}

\subsubsection{Backgrounds that respect SSA}

\begin{figure*}[h]
$$
\begin{array}{cc}
  \includegraphics[width=7cm]{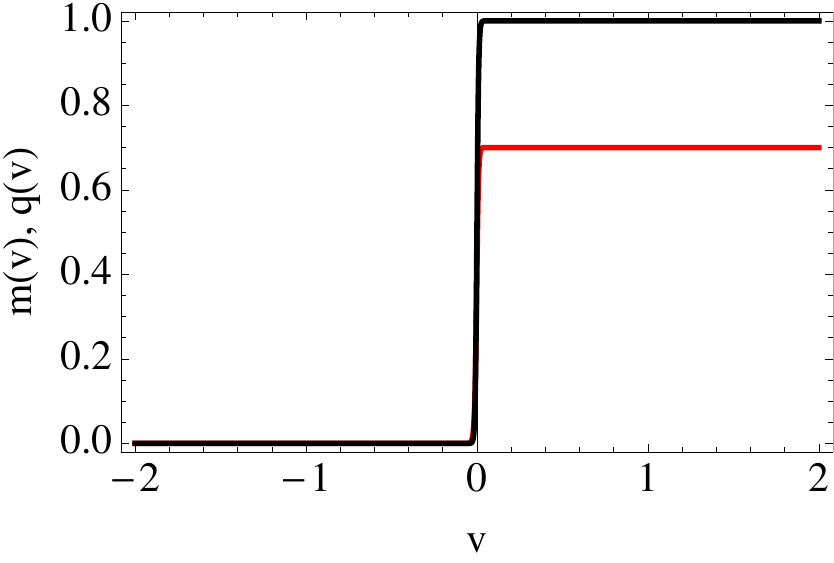} & \includegraphics[width=7cm]{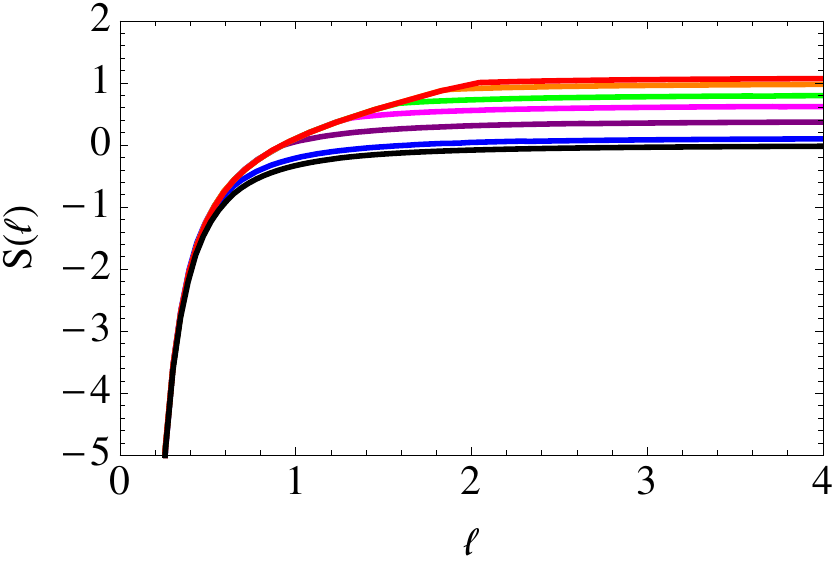}
\end{array}
$$
\caption{Left panel: $m(v)$ (black) and $q(v)$ (red). Right panel: $S(\ell)$, for $t_b=$ 0.01 (black), 0.5 (blue), 1 (purple), 1.5 (magenta), 2 (green), 3 (orange) and 5 (red).} \label{4dchoice1}
\end{figure*}

\begin{figure*}[h]
$$
\begin{array}{cc}
  \includegraphics[width=7cm]{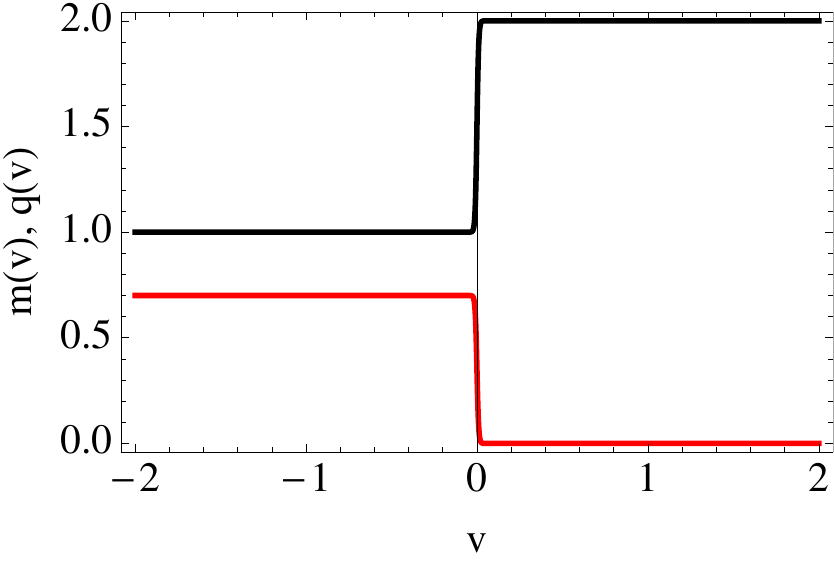} & \includegraphics[width=7cm]{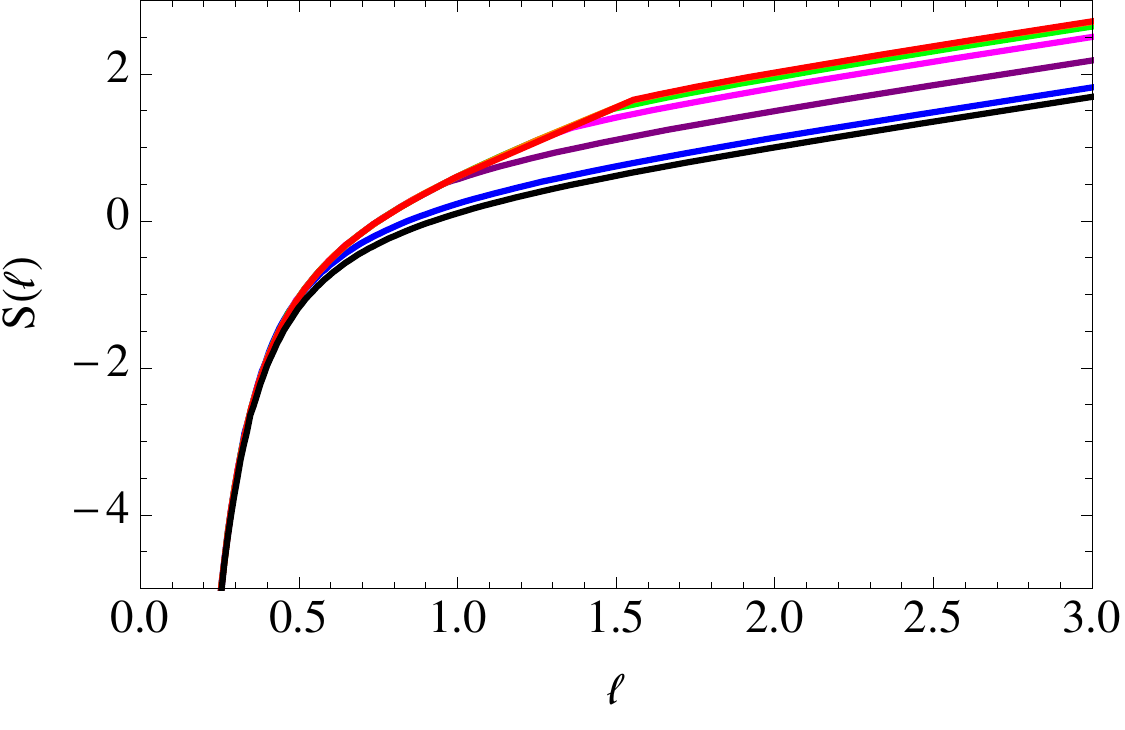}
\end{array}
$$
\caption{Left panel: $m(v)$ (black) and $q(v)$ (red). Right panel: $S(\ell)$, for $t_b=$ 0.01 (black), 0.5 (blue), 1 (purple), 1.5 (magenta), 2 (green), 3 (orange) and 5 (red).}\label{4dchoice2}
\end{figure*}

We will discuss similar processes as in the case for $d=3$. The qualitative features are very similar here, and hence we will limit ourselves in terms of the details. The analogous choices that preserve SSA are pictorially represented in Figure \ref{4dchoice1} and \ref{4dchoice2}. Note that, Figure \ref{4dchoice1} corresponds to the choices used in \cite{Caceres:2012em} to analyze the scaling of the thermalization time with temperature and chemical potential of the system. 

\begin{figure*}[h]
$$
\begin{array}{cc}
  \includegraphics[width=7cm]{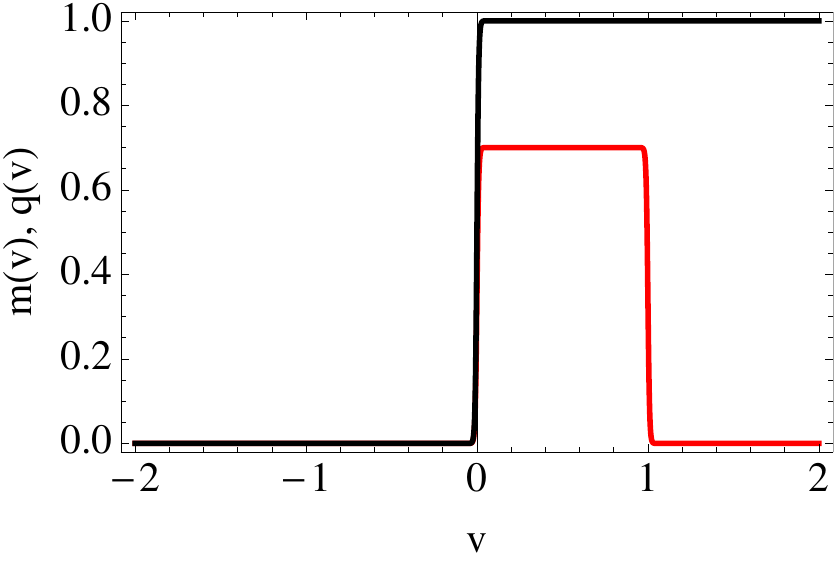} & \includegraphics[width=7cm]{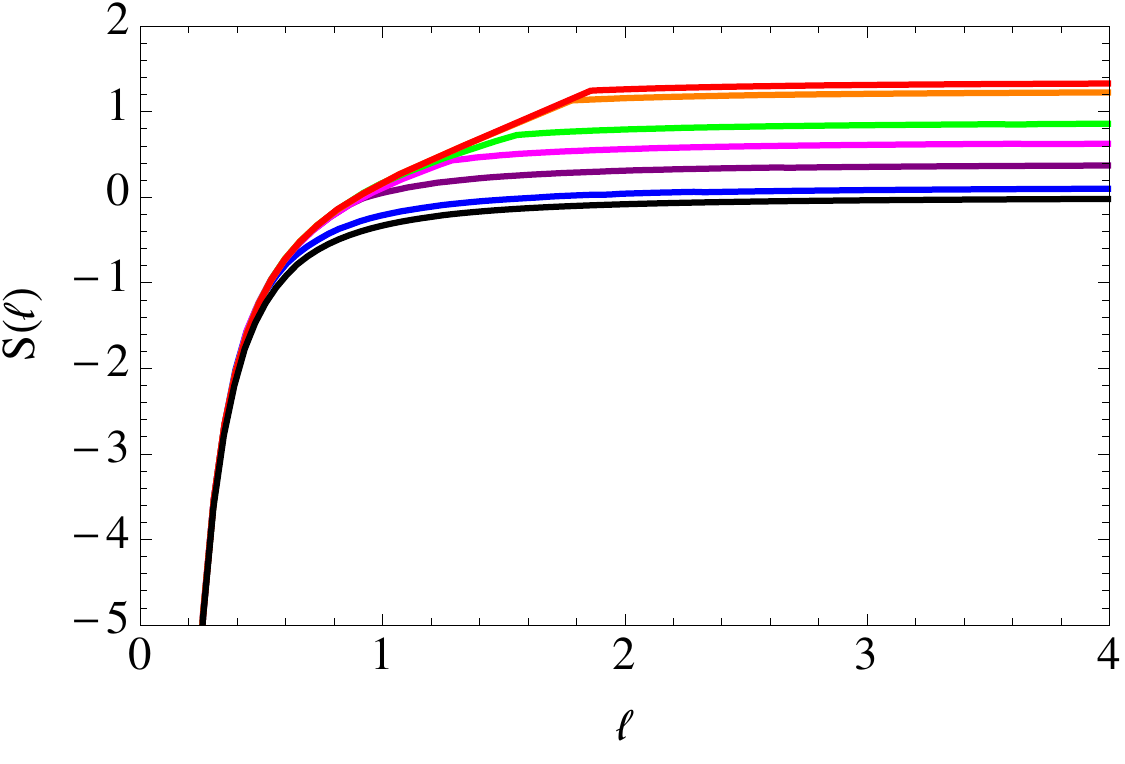}
\end{array}
$$
\caption{Left panel: $m(v)$ (black) and $q(v)$ (red). Right panel: $S(\ell)$, for $t_b=$ 0.01 (black), 0.5 (blue), 1 (purple), 1.5 (magenta), 2 (green), 3 (orange) and 5 (red).} \label{4dchoice3}
\end{figure*}

%

\subsubsection{Backgrounds that violate SSA}

\begin{figure*}[h]
$$
\begin{array}{cc}
  \includegraphics[width=7cm]{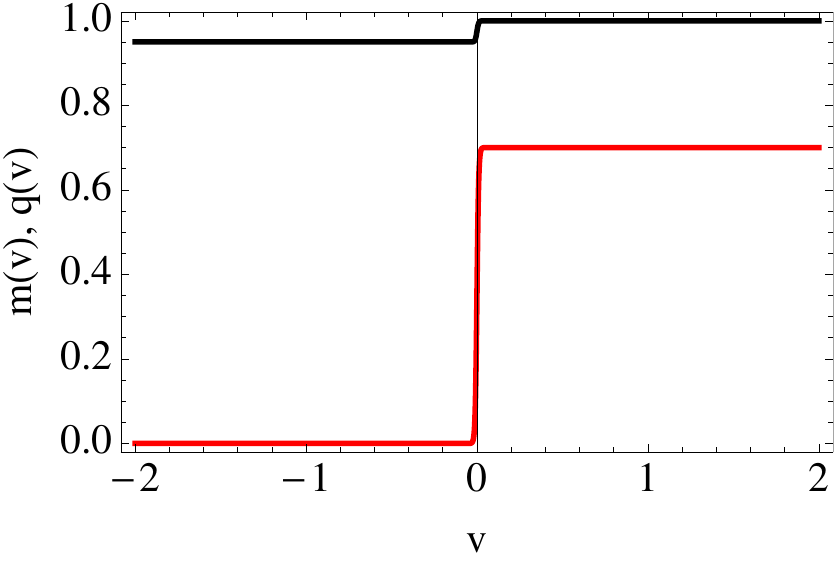} & \includegraphics[width=7cm]{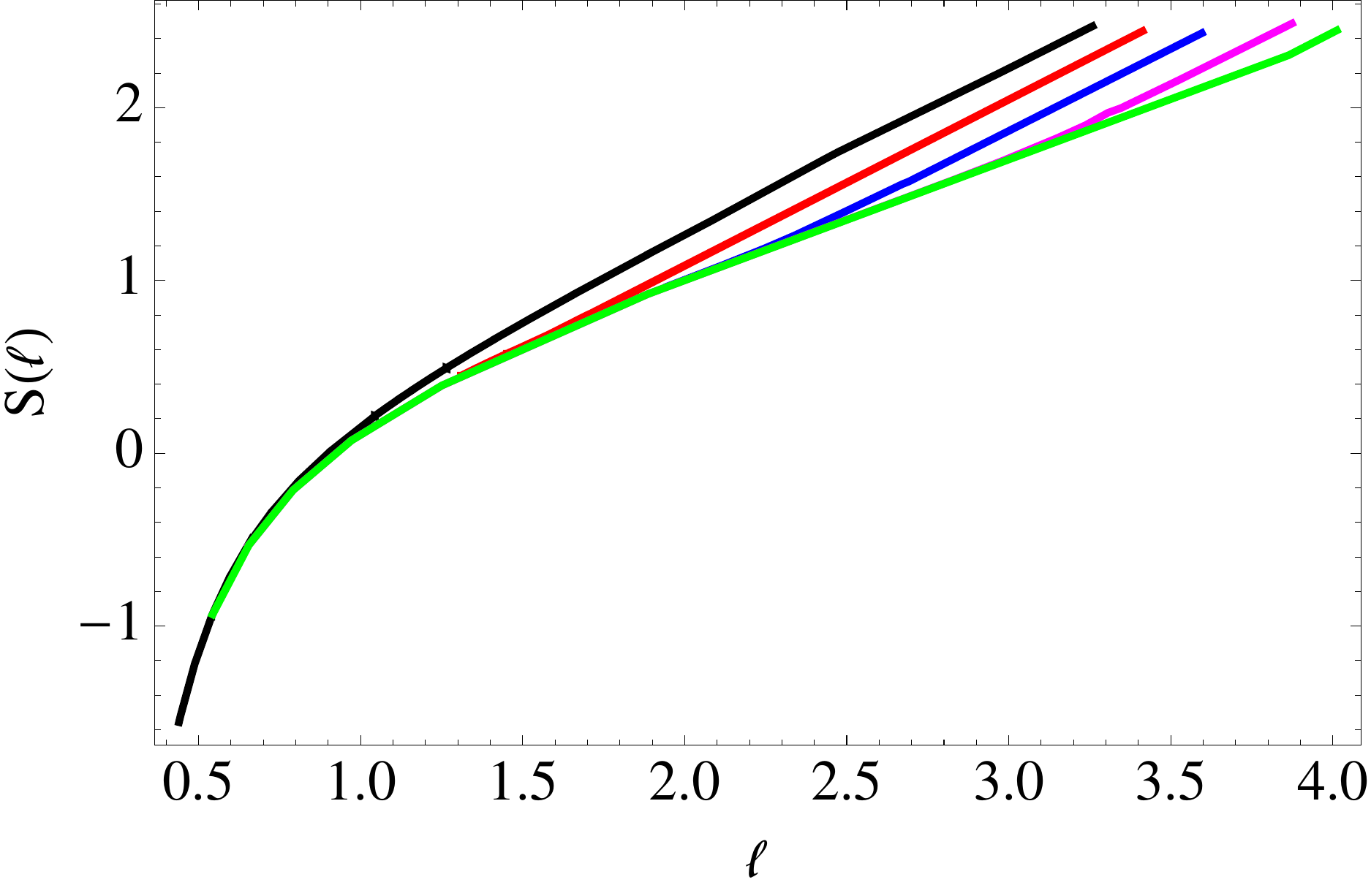}
\end{array}
$$
\caption{The mass and charge functions for which SSA violation occurs: $m(v)$ (black) and $q(v)$ (red). Right panel: $S(l)$ for $t_b=$ $0.5$ (black), $1.5$ (red), $2.0$ (blue), $2.5$ (magenta) and  $3.0$ (green)} \label{4dchoicebad1}
\end{figure*}

\begin{figure*}[h]
$$
\begin{array}{cc}
  \includegraphics[width=7cm]{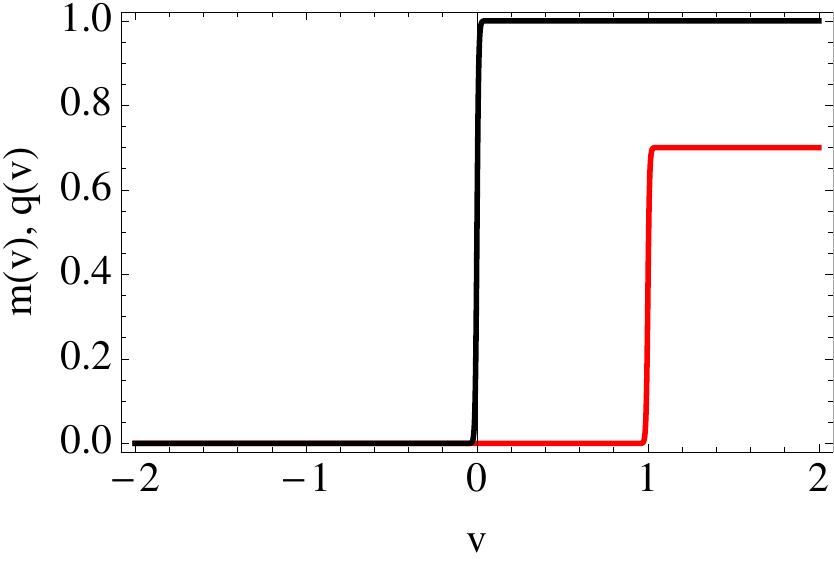} & \includegraphics[width=7cm]{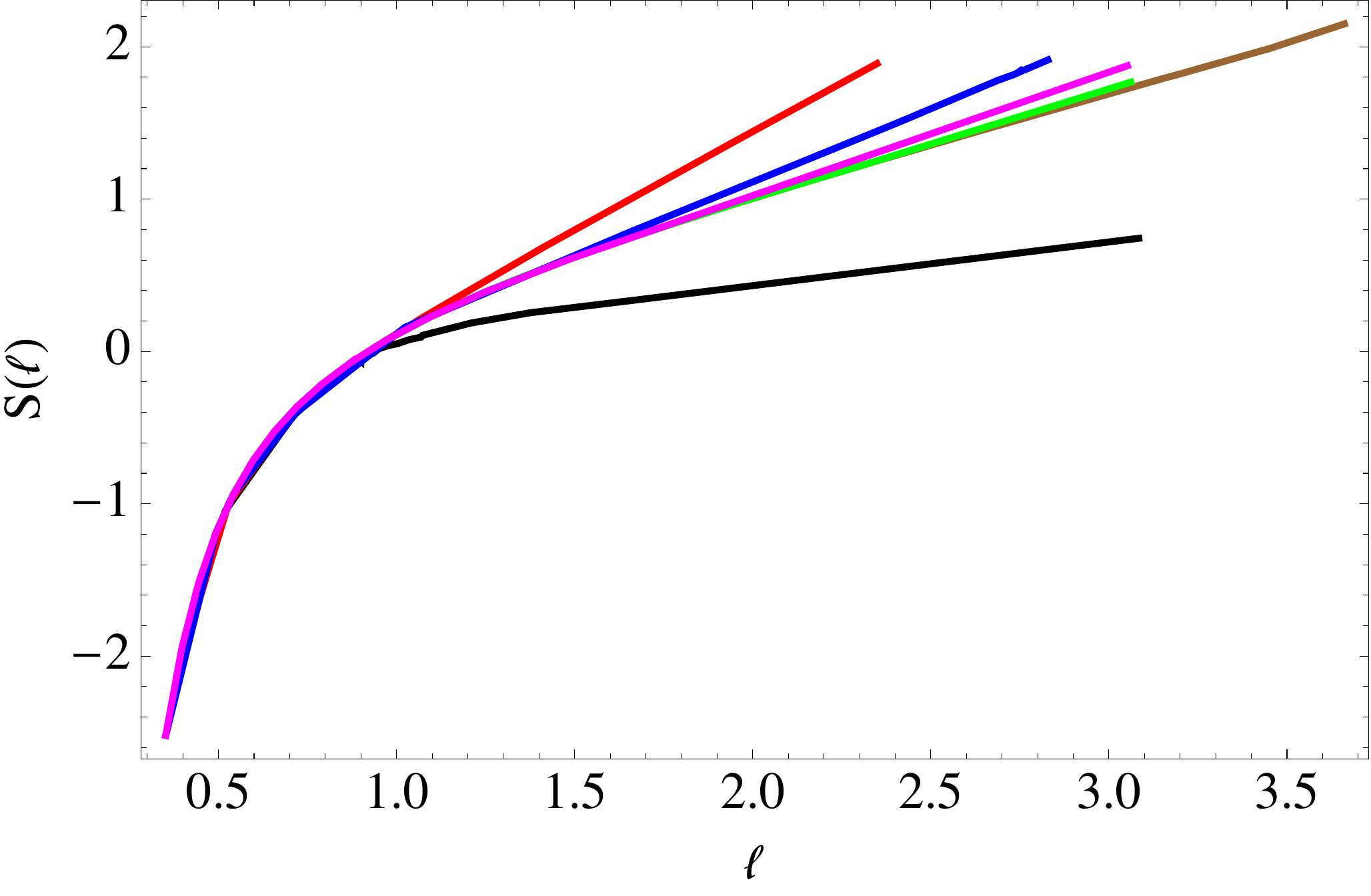}
\end{array}
$$
\caption{The mass and charge functions for which SSA violation occurs: $m(v)$ (black) and $q(v)$ (red). Right panel: $S(l)$ for $t_b= 1.0$ (black), $1.5$ (red), $2.5$ (blue),$3.0$ (magenta) and $ 3.5$ (green). For the blue and magenta curves we have numerically verified the change in concavity. } \label{4dchoicebad2}
\end{figure*}

Once more, we will keep our discussion very brief and present the examples that violate SSA. These are shown in Figure \ref{4dchoicebad1} and Figure \ref{4dchoicebad2}). The choices are analogous to the ones made in (\ref{exam4}) and (\ref{exam5}. In  Figure \ref{4dchoicebad2}  the change in concavity is not as  clear from the graph as in the other examples but it is easy to verify it numerically   \footnote{Since  the curves are initially concave we just have to verify that after certain value of $l=l_0$  they become convex. Namely, we verify that if we  take any two points $x_1, x_2 > l_0$  and we will have   $S(y x_1 + (1-y) x_2) \le y S(x_1) + (1-y) S(x_2)$ where $0<y<1$. }.  This physics can also be observed by studying the geodesics, which penetrate the critical surface whenever there is a violation of SSA but not otherwise.
%

\section{Discussion and conclusions}

In this article we have explored and demonstrated an interesting pictorial realization of the strong sub-additivity condition in terms of the bulk gravitational description. In the presence of charge, the dynamical evolution from a low temperature, low chemical potential pure state to a thermalized state with a non-zero value of the chemical potential does indeed sharpen the connection between the bulk null energy condition constraint and the strong sub-additivity of entanglement entropy in the boundary field theory, which was alluded to in \cite{Allais:2011ys, Callan:2012ip}.

Our investigations suggest that the dual field theory disallows specific choices of the mass and the charge functions for which it is possible to penetrate the critical surface. However, as we have learned now, the nature of the violation of the NEC depends on the class of examples we choose; such as the details which have qualitatively distinct behaviours for the charged case and the uncharged one. Generally, the NEC is an algebraic constraint on the bulk energy-momentum tensor, which --- should a general result exist --- may correspond to an algebraic constraint in the boundary theory as well. Strong sub-additivity depends crucially on the concavity property of the entropy function and thus it is an intriguing possibility to consider establishing a direct {\it equivalence} between the bulk null energy condition and the concavity property of the entropy function at the boundary. See \cite{Wall:2012uf} for some recent progress towards a {\it proof}; however, it does not necessarily apply for backgrounds where a black hole eventually forms.

Coming back to our case, it is perhaps surprising how SSA can be obeyed for some examples, specially since the critical surface always exists for any generic mass and charge functions. Unlike the time-independent cases, where no extremal surface can penetrate the black hole event horizon, the Vaidya backgrounds give rise to an apparent horizon that can be penetrated by a space-like surface. The critical surface may lie above or below this apparent horizon. If it is below ($z_c < z_{\rm ah}$)  there is violation of NEC and SSA; If it is above ($z_c > z_{\rm ah}$)  NEC and SSA are respected. Let us emphasize that there is no {\it a priori} criterion that prohibits the minimal surface to penetrate the critical surface when it is cloaked by the apparent horizon. Nevertheless, this is what we observe. It would be interesting to understand how general this feature is.  We can also ask   if the analysis changes if  we consider  extremal black holes. If the initial state is the vacuum and the final state is extremal (mass and charge functions similar to Figure \ref{4dchoice1}), the critical surface is cloaked by the apparent horizon ($z_c > z_{\rm ah}$)  and NEC and SSA are  obeyed, in agreement with our observations. On the other hand, if  the initial state is extremal and the final state is an arbitrary thermal state we cannot conclude something general. However, we do not see  new features emerging in the analysis.

Therefore, based on our  observations we can venture a {\it na\'{i}ve} characterization for the choice functions. Let us assume that $m'(v) \ge 0$. This is required by continuity with the results known in $q(v) \to 0$ limit. Given this, we can characterize a charge function $q(v)$ to be {\it good} if $z_c > z_{\rm ah}$ for all times and {\it bad} if $z_c < z_{\rm ah}$ for any time. Of course, we also need to impose a constraint on the maximum magnitude of the mass and the charge functions in order to avoid the {\it naked}. Such a characterization, at present, is only a plausibility.

It is intriguing that the SSA condition seems to constrain the global time-evolution process but does not say anything about the initial conditions. In general, it is possible that given an {\it arbitrary but reasonable} initial condition the dual field theory undergoes time-evolution, but never thermalizes or obtains a steady-state phase. Such a process, once obtained by solving Einstein gravity with a reasonable matter field in the bulk, will surely preserve NEC and hence SSA. It will thus be interesting to investigate whether the SSA condition plays a similar role for more conventional systems rather then large $N$ gauge theories having a gravity dual. We hope to address these issues in details in future.

\section{Acknowledgements}

We would like to thank Sumit Das and Veronika Hubeny for helpful discussions. E.C. acknowledges support of CONACyT grant CB-2008-01-104649. This  material is based upon work supported by the National Science Foundation under Grant PHY-0969020 and by the Texas Cosmology Center. AK is supported by the Simons postdoctoral fellowship awarded by the Simons Foundation. WT is also supported by a University of Texas continuing fellowship.


\end{document}